\documentclass[reprint,amsmath,amssymb,aps,floatfix,showkeys,showpacs,
]{revtex4-1}

\pdfoutput=1 
\usepackage[utf8]{inputenc}
\usepackage[T1]{fontenc}
\usepackage{amsmath,amssymb}
\usepackage{graphicx,colordvi,longtable,color}
\usepackage{dcolumn}
\usepackage{bm}

\newcommand{\cn}{\,{\sf cn}}

\newcommand{\dn}{\,{\sf dn}}
\newcommand{\sech}{\,{\sf sech}}

\newcommand{\tnh}{\,{\sf tanh}}

\begin{document}


\title{Generalized KdV-type equations versus Boussinesq's equations for uneven bottom - numerical study}


\author{Anna Karczewska}
 \email{A.Karczewska@wmie.uz.zgora.pl}
\affiliation{Faculty of Mathematics, Computer Science and Econometrics\\ University of Zielona G\'ora, Szafrana 4a, 65-246 Zielona G\'ora, Poland}

\author{Piotr Rozmej}
 \email{P.Rozmej@if.uz.zgora.pl}
\affiliation{Faculty of Physics and Astronomy,
University of Zielona G\'ora, Szafrana 4a, 65-246 Zielona G\'ora, Poland}

\date{\today}

\begin{abstract} 
The paper's main goal is to compare the motion of solitary surface waves resulting from two similar but slightly different approaches. In the first approach, the numerical evolution of soliton surface waves moving over the uneven bottom is obtained using single wave equations. In the second approach, the numerical evolution of the same initial conditions is obtained by the solution of a coupled set of the Boussinesq equations for the same Euler equations system. We discuss four physically relevant cases of relationships between small parameters $\alpha,\beta,\delta$. For the flat bottom, these cases imply the Korteweg-de Vries equation (KdV), the extended KdV (KdV2), fifth-order KdV (KdV5), and the Gardner equation (GE). In all studied cases, the influence of the bottom variations on the amplitude and velocity of a surface wave calculated from the Boussinesq equations is substantially more significant than that obtained from single wave equations.
\end{abstract}

\pacs{ 02.30.Jr, 05.45.-a, 47.35.Bb, 47.35.Fg}
\keywords{KdV-type equations, Gardner equation, uneven bottom, numerical evolution }

\maketitle



\section{Introduction - the concept of the study} \label{idea}

Nonlinear waves are the subject of a vast number of studies in many fields of science. They appear in hydrodynamics, propagation of optical and acoustic waves, plasma physics, electrical circuits, biology, and many others. These equations usually appear as approximations of more basic laws describing the behavior of relevant systems, usually too complicated for non-numerical analysis. These approximations assume that some parameters characterizing the system are small, and then a perturbative approach can be used. In this way, one can derive various nonlinear wave equations, e.g., the \emph{Korteweg-de Vries equation} (KdV), the \emph{extended Korteweg-de Vries equation} (KdV2), \emph{5th-order KdV} or the \emph{Gardner} equation. All these equations can be derived from the Euler equations describing the model of the irrotational motion of an inviscid and incompressible fluid in a container with a flat, impenetrable bottom.

The real world, however, is not that simple. In particular, bottoms of oceans, seas, rivers are non-flat. Therefore, it would be desirable to find a relatively simple mathematical description that would take into account bottom variations. In the past, there were many attempts to attack this problem. In this article, we only briefly remind some of these works. Some first results were obtained by Mei and Le M\'ehaut\'e \cite{Mei}, and Grimshaw \cite{Grim70}. Several authors \cite{Djord,BH} studied these problems using variable coefficient nonlinear Schr\"odinger equation (NLS). Some research groups developed approaches combining linear and nonlinear theories \cite{Pel,Peli,Peli1}. The Gardner equation  was also extensively investigated in this context \cite{Grim,Smy,Kam,PS98}. 
The Hamiltonian approach was utilized by Van Groeasen and Pudjaprasetya \cite{G&P1,G&P2}.
Another widely applied method consists in taking an appropriate average of vertical variables, which results in the Green-Naghdi equations \cite{GN,Nad,Kim}. Several authors derived variable coefficient KdV equation (vcKdV) \cite{MM69,Kak71,John72,John73,Ben92} in attempts to describe the evolution of a solitary wave moving onto a shelf.
Article \cite{RoPa83} is the only one known to us  (apart from our approach) in which the authors introduce besides two small standard parameters, the third one associated with an uneven bottom. We presented a broader discussion of some of the current attempts and methods to account for uneven bottoms in \cite{KRcnsns}.

In the paper \cite{KRcnsns}, we derived equations of the KdV type for an uneven bottom for various relationships between small parameters $ \alpha, \beta, \delta $. For a flat bottom, one can always eliminate the $ w $ function from the Boussinesq equations and get a single wave equation for the $ \eta $ function (surface distortion from the equilibrium state). For an uneven bottom, this can only be done for the lowest possible order of the perturbation approach, and only if the bottom is a piecewise linear function. In other cases, there is no $ w $ function that makes the Boussinesq equations compatible. Therefore, for testing surface waves in the case of an uneven bottom studying the set of Boussinesq's equation seems to be more appropriate.  The present work supplements \cite{KRcnsns} with a comparison of these two methods, including the study of the Gardner equation and calculations for much longer evolution times.

In \cite{KRcnsns}, we derived four new wave equations, which generalize for the case of  uneven bottom the Korteweg-de Vries equation (KdV), the extended KdV (KdV2), the fifth-order KdV, and the Gardner equation (combined KdV - mKdV). The first is obtained for $\alpha=O(\beta)$, $\delta=O(\beta)$, the second for $\alpha=O(\beta)$, $\delta=O(\beta^2)$, the third for $\alpha=O(\beta^2)$, $\delta=O(\beta^2)$ and the fourth for  $\beta=O(\alpha^2)$, $\delta=O(\beta^2)$. In all cases, the generalized wave equations could be derived only for a particular class of bottom functions, namely the piecewise linear ones. On the way to these results, we derived corresponding sets of the Boussinesq equations, which are valid for bottoms of arbitrary shapes. 

However, it seems that in numerical simulations of wave evolution according to these generalized equations, all of them can be used for arbitrary bottom functions. The reason consists in the discretization of numerical codes. The knowledge of the bottom function is needed only in the mesh points, like when the bottom function is a piecewise linear one.

In the paper, we numerically test the results of the evolution of the nonlinear waves obtained from the Boussinesq equations with those obtained from the corresponding single KdV-type equations generalized for the uneven bottom in \cite{KRcnsns}.  We assume that initial conditions correspond to solitons appropriate to the particular case. Such soliton can be formed in a region of flat bottom, and next enter the region where the bottom is varying. 

The paper is organized as follows. In section \ref{EulEq} we briefly remind the reader of the Euler equations for the irrotational motion of the inviscid, incompressible fluid, which arises for the shallow water problem. This set of equations can serve as a starting point for the derivation of both Boussinesq's equations and the single wave equation for each particular case of ordering of small parameters. In section \ref{a1b1d1} the case of generalized KdV equation is analyzed. In section \ref{2bus} we discuss the generalized extended KdV (KdV2). Next, in section \ref{5bus} the generalized fifth-order KdV is studied. Section \ref{GE} is devoted to the generalized Gardner equation. In section \ref{nonSol}, we studied some examples in which the initial conditions are substantially different from the solitons appropriate for particular equations.
The conclusions are contained in Section \ref{concl}.

\section{Euler equations for an uneven bottom} \label{EulEq}

To make the paper self-contained, we briefly remind the approach to the shallow water problem in a more general case when the bottom of the fluid is not even. The model applies to the waves on both the surface of the liquid and the interface between two immiscible fluids.  A detailed description of the model and methods of deriving relevant nonlinear wave equations is presented in our work \cite{KRcnsns}. 

The set of Euler equations, written in nondimensional variables has the following form  
\begin{align} \label{2BS}
\beta \phi_{xx}+\phi_{zz} & = 0, \\ \label{4BS}
\eta_t+\alpha\phi_x\eta_x-\frac{1}{\beta}\phi_z & = 0,
\\  \label{5BS}
\phi_t+\frac{1}{2}\alpha \phi_x^2+\frac{1}{2}\frac{\alpha}{\beta}\phi_z^2 +\eta 
 -\tau \beta \frac{\eta_{2x}}{(1+\alpha^2\beta\eta_x^2)^{3/2} }
& = 0,
\\ \label{6BS}
\phi_z-\beta\delta\left( h_x\,\phi_x\right) & =  0.
\end{align}
Equation (\ref{2BS}) is the Laplace equation for the velocity potential valid for the whole volume of the fluid. Equations (\ref{4BS}) and (\ref{5BS}) are so-called kinematic and dynamic boundary conditions at the surface, that is for $z = 1+\alpha \eta$, respectively. The equation  (\ref{6BS}) represents the boundary condition at the non-flat unpenetrable bottom, i.e.~for $z=\delta h(x)$. 
In (\ref{5BS}), the Bond number $\tau=\frac{T}{\varrho g h^2}$, where $T$ is the surface tension coefficient. For surface gravity waves, this term can be safely neglected, since $\tau < 10^{-7}$ (when the fluid depth is of the order of meters), but it can be important for waves in thin fluid layers.
For abbreviation all subscripts in (\ref{2BS})-(\ref{6BS}) denote the partial derivatives with respect to particular variables, i.e.\ $\phi_{t}\equiv\frac{\partial \phi}{\partial t}, \eta_{2x}\equiv\frac{\partial^2 \eta}{\partial x^2}$, and so on. 

The parameters $\alpha,\beta,\delta$ in the set (\ref{2BS})-(\ref{6BS}) have the following meaning. Besides standard small parameters $\alpha=\frac{a}{H}$ and $\beta=\left(\frac{H}{l}\right)^2$ we introduced the third one, defined as $\delta=\frac{a_h}{H}$. Here $a$ represents the wave amplitude, $H$ - average depth of the basin, $l$ - average wavelength and $a_h$ - amplitude of bottom variations. For the perturbation approach, all of them should be small, however not necessarily of the same order. Therefore for different ordering of these parameters one can derive different sets of the Boussinesq equations and in consequence different wave equations. The cases with flat bottom $(\delta=0)$ are presented in \cite{BurSerg}. We already introduced the third small parameter $\delta=\frac{a_h}{H}$ in \cite{KRI} in order to generalize the extended KdV equation (KdV2) for the case of the uneven bottom. Unfortunately, the derivation presented in \cite{KRI} is not fully consistent, and the final equation contains an improper term additionally.
  
As usual, the velocity potential is seeking in the form of power series in the vertical coordinate
\begin{equation} \label{Szer}
\phi(x,z,t)=\sum_{m=0}^\infty z^m\, \phi^{(m)} (x,t),
\end{equation}
where ~$\phi^{(m)} (x,t)$ are yet unknown functions. The Laplace equation (\ref{2BS}) determines $\phi$ in the form, which involves only two unknown functions with the lowest $m$-indexes, $f(x,t):=\phi^{(0)} (x,t)$ and $F(x,t):=\phi^{(1)} (x,t)$. Hence,
\begin{align} \label{Szer1}
\phi(x,z,t) & =\sum_{m=0}^\infty \frac{(-1)^m\beta^m}{(2m)!} \frac{\partial^{2m} f(x,t)}{\partial x^{2m}} z^{2m} \\ & 
 + \sum_{m=0}^\infty \frac{(-1)^m\beta^{m+1}}{(2m+1)!} \frac{\partial^{2m+1} F(x,t)}{\partial x^{2m+1}} z^{2m+1}. \nonumber
\end{align}
The explicit form of this velocity potential reads as
\begin{align} \label{pot8}
\phi = f & -\frac{1}{2}\beta z^2 f_{2x} + \frac{1}{24}\beta^2 z^4 f_{4x} - \frac{1}{720}\beta^3 z^6 f_{6x} + \cdots  \nonumber \\ & 
+ \beta z F_x -\frac{1}{6}\beta^2 z^3 F_{3x}+ \frac{1}{120}\beta^3 z^5 F_{5x}
+ \cdots 
\end{align}

In the next step, one uses the boundary condition at the bottom (\ref{6BS}). For a standard flat bottom case it follows that $F_x=0$ and only $f$ and its even $x$-derivatives remain in (\ref{pot8}). For an uneven bottom, the situation is more complicated, and one can express $F_x$ explicitly by $f$ only in some low order. Precisely this order depends on the relation between $ \beta$ and $\delta$ parameters. Below we show this step explicitly for the case $\delta=O(\beta)$. For other cases, in which the procedure is analogous, śwe refer to \cite{KRcnsns}.
Insertion of the velocity potential (\ref{pot8}) into (\ref{6BS}) gives (with $z=\delta h(x)$) the following complicated relation between the functions $F_x$ ~and~$f$ 
\begin{align} \label{fF}
F_{x} -\delta (h f_x)_x & -\frac{1}{2} \beta\delta^2 (h^2 F_{2x})_x +\frac{1}{6} \beta\delta^3 (h^3 f_{3x})_x \nonumber \\ &+\frac{1}{24}\beta^2 \delta^4 (h^4 F_{4x})_x+ \cdots=0.
\end{align}
Keeping only terms lower than third order leaves 
\begin{equation} \label{F2}
F_x =\delta (h f_x)_x,
\end{equation}
which allows us to express the $x$-dependence of the velocity potential through $f,h$, and their $x$-derivatives up to second order. This fact limits the velocity potential to the form
\begin{align} \label{p8}
\phi = f & -\frac{1}{2}\beta z^2 f_{2x} + \frac{1}{24}\beta^2 z^4 f_{4x} 
+ \beta\delta z (h f_x)_x
\end{align}
valid only up to second order in small parameters. 
Attempts to go to higher orders would require solving the equation (\ref{fF}) for $F$ with arbitrary $h$, which is impossible to do.

\section{Case $\alpha=O(\beta)$, ~$\delta=O(\beta)$ - generalization of KdV
} \label{a1b1d1}

This case corresponds to shallow water waves. Since the coefficient of surface tension is very small, one can safely neglect the appropriate term in the Euler equations.

\begin{figure}[htb] \begin{center}
 \resizebox{0.99\columnwidth}{!}{\includegraphics{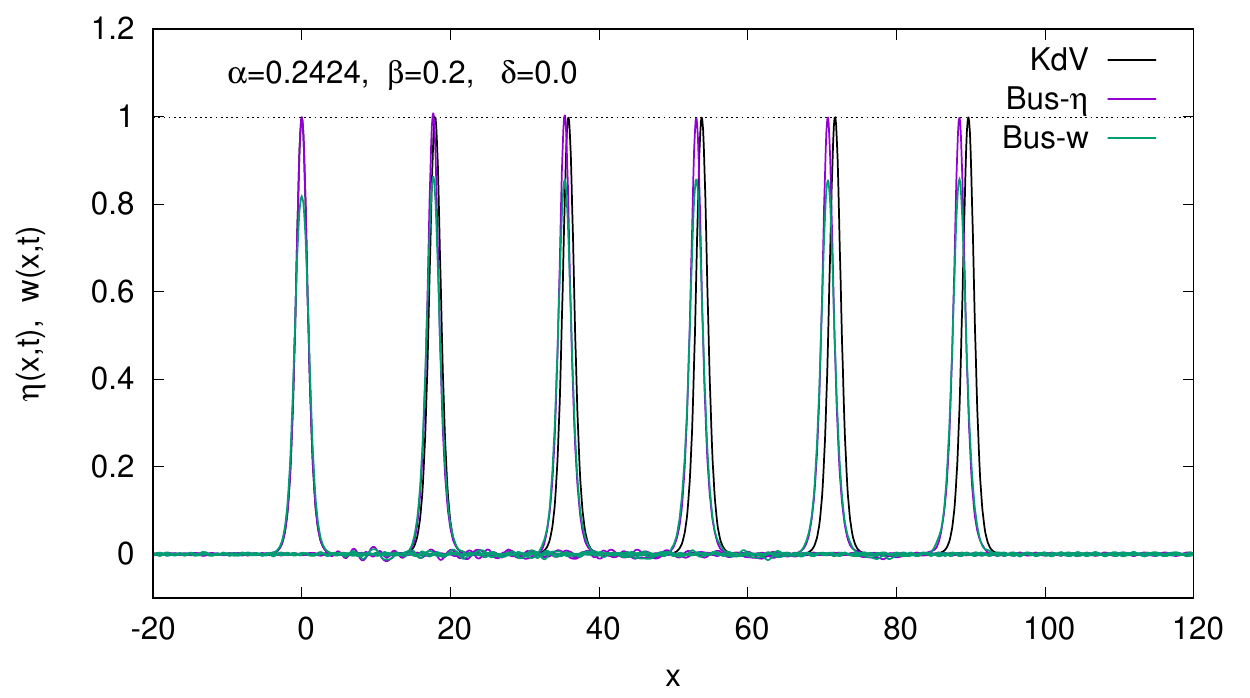}}
 \caption{Time evolution of the KdV soliton (\ref{1Sol}) obtained according to KdV equation (\ref{kdvD}) -- black lines and that obtained from the Boussinesq set (\ref{4hx1})-(\ref{5hx1}) -- blue lines. Additionally, the evolution of $w(x,t)$ function is displayed with green lines. Flat bottom ($\delta=0$) is assumed.}  \label{a24b2d0K.}
\end{center} \end{figure}

Due to the presence of the term $-\frac{1}{\beta}\phi_z$ in (\ref{4BS}), the Boussinesq equations resulting from the substitution of (\ref{p8}) into (\ref{4BS}) and (\ref{5BS}) are correct only up to first order in $\alpha, \beta$ and $\delta$.
They take the following form (see, \cite{KRcnsns}, eqs.\ (17)-(18))
\begin{align} \label{4hx1}
  \eta_t + w_x & + \alpha (\eta w)_x-\frac{1}{6}\beta w_{3x}  -
 \delta (hw)_x =0,  \\ \label{5hx1}
w_t + \eta_x & + \alpha w w_x -\frac{1}{2}\beta  w_{2xt} =0 . 
\end{align} 
Elimination of $w$ from (\ref{4hx1})-(\ref{5hx1}) in order to obtain a single wave equation for $\eta$ appears to be possible only when $h_{2x}=0$, that is when the bottom function is the piecewise linear one. In such case the system (\ref{4hx1})-(\ref{5hx1}) can be  made compatible, and reduced to the single KdV-type equation (\cite{KRcnsns}, eq.\ (28))
\begin{equation} \label{kdvD} 
\eta_t + \eta_x + \frac{3}{2} \alpha \eta\eta_x+\frac{1}{6}\beta \eta_{3x} - \frac{1}{4} \delta (2 h \eta_x+h_x \eta)  =0.
\end{equation}
On the other hand, the Boussinesq equations do not require the condition $h_{2x}=0$, the bottom function $h$ can be arbitrary. From this point of view the Boussinesq equations (\ref{4hx1})-(\ref{5hx1}) are more general (more fundamental) than the single wave equation (\ref{kdvD}).

It is worth to emphasize that the above properties are general. They are the same for all cases (all wave equations) discussed in this paper. For more details on the derivation of nonlinear wave equations generalized for the uneven bottom, we refer to \cite{KRcnsns}.

\begin{figure}[htb] \begin{center}
\resizebox{0.9\columnwidth}{!}{\includegraphics{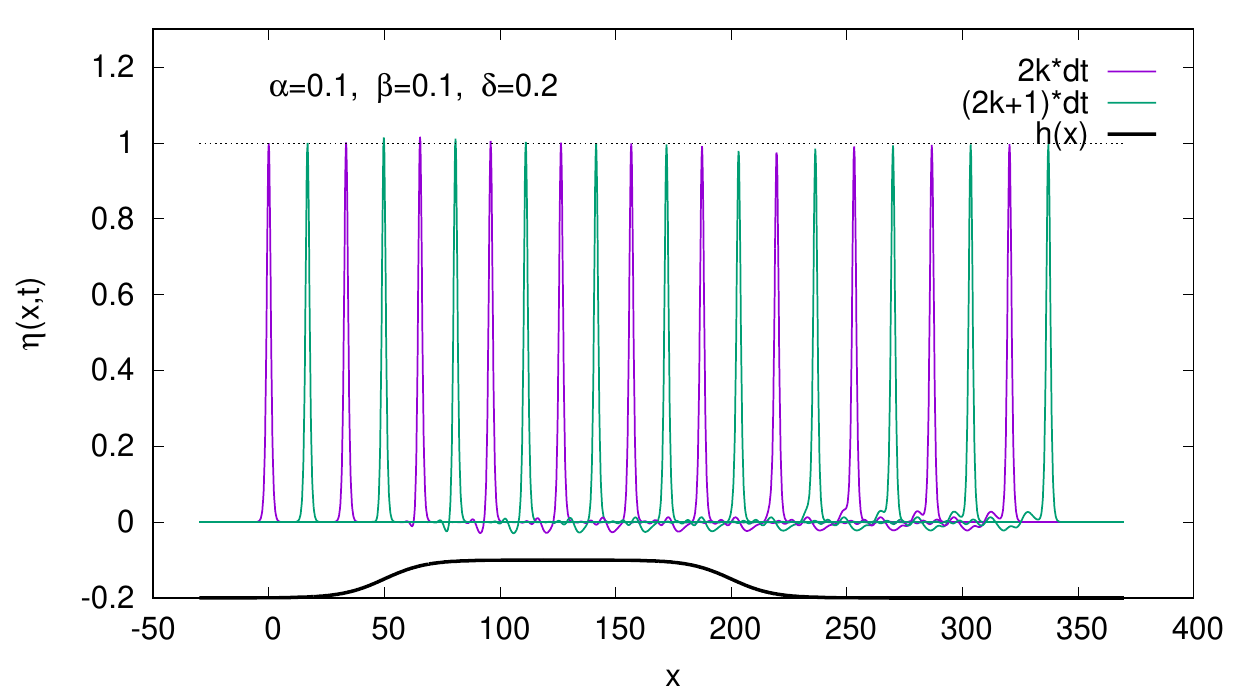}} \caption{Time evolution of the KdV soliton (\ref{1Sol}) obtained according to KdV equation (\ref{kdvD}) for the bottom given by (\ref{Btanh}) with $\delta=0.2$. Subsequent profiles correspond to times $t_n= n*16$, with $n=0,1,\ldots,21$.
 The shape of the bottom function is drawn in an arbitrary scale.} \label{a1b1d2KdV.}
\end{center} \end{figure}

In numerical simulations, we can apply the FDM (finite difference method) with leap-frog, which stability is well determined for appropriate relation between time step $\Delta t$ and mesh size $\Delta x$.

For the equation (\ref{kdvD}) the appropriate algorithm is the following
\begin{align}\label{dkdv} 
\eta_i^{j+1} & = \eta_i^{j-1} -2\Delta t \left((\eta_x)_i^{j-1}+\frac{3}{2}\alpha \eta_i^{j-1} (\eta_x)_i^{j-1} \right.\\ & \hspace{2ex}\left. + \frac{1}{6} \beta(\eta_{3x})_i^{j-1} -\frac{1}{4} (2 h_i (\eta_x)_i^{j-1} +(h_x)_i \eta_i^{j-1} )\right) . \nonumber 
\end{align}

For the Boussinesq set (\ref{4hx1})-(\ref{5hx1}), we have to evolve two equations simultaneously
\begin{align} \label{deta}
\eta_i^{j\!+\!1} & =  \eta_i^{j\!-\!1}\! -\!2\Delta t \Big[ (w_x)_i^{j\!-1\!} \!+\!\alpha\left((\eta_x)_i^{j\!-\!1}w_i^{j\!-\!1}\!+\! \eta_i^{j\!-\!1} (w_x)_i^{j\!-\!1}\right) \nonumber  \\ & - \frac{1}{6}\beta(w_{3x})_i^{j-1} 
-\delta \left( (h_x)_i w_i^{j-1} +h_i(w_x)_i^{j-1}\right)\Big] \\  \label{dw}
w_i^{j\!+\!1} = & w_i^{j\!-\!1} \!-\! 2\Delta t \left((\eta_x)_i^{j\!-\!1} +\alpha w_i^{j\!-\!1}(w_x)_i^{j\!-\!1} - \frac{1}{2} \beta (w_{2xt})_i^{j\!-\!1} \right) .
\end{align} 

In (\ref{dkdv})-(\ref{dw}) ,
 $i=0,1\ldots,N-1$ is the index of the mesh point $x_i$ and $j$ enumerates time step. Periodic boundary conditions in $x$ are used. Time increment $\Delta t=\frac{(\Delta x)^3}{4}$ assures stability of the time integration. 
Setting $\delta=0$ one obtains the set of equations corresponding to the Korteweg-de Vries equation.

In first tests of the code we use initial condition in the form of the KdV soliton, that is, $\eta(x,0)$, where
\begin{align}\label{1Sol}
\eta(x,t)&=A\,\text{\sech}^2\left[ \sqrt{\frac{3\,\alpha}{4\,\beta }A\,}  \left(x- t \left(1+A\frac{\alpha }{2}\right)\right)\right] \nonumber \\ &
=A\,\text{\sech}^2\left[ B(x-vt)\right].
\end{align}
Then the initial condition for $w$ is given by 
$$ w= \eta-\frac{1}{4}\alpha \eta^2+ \frac{1}{3}\beta \eta_{2x} \quad \mbox{at}\quad t=0.$$ 

In Fig.~\ref{a24b2d0K.}, numerical 
results of the KdV soliton (\ref{1Sol}) evolution for $\alpha=0.2424$, $\beta=0.2$ and $\delta=0$, that is for the flat bottom, are shown. The KdV soliton amplitude is chosen to be $A=1$ for comparison with the KdV2 case shown in Fig.\ \ref{a24b2BUS.}. In both Figs. 1 and 3, time separation between displayed wave profiles is $dt=16$. Results, shown in Fig.~\ref{a24b2d0K.}, can be considered as a check of the numerical code. In the KdV case, the soliton moves with the constant velocity ($v=1+\frac{\alpha}{2}A)$ and a fixed profile. Since initial conditions are chosen as KdV soliton, the $\eta$ and $w$ functions evolving according to Boussinesq's equations develop very small tails and move with slightly different velocity, but profiles of their main parts exhibit a soliton motion.

\begin{figure}[hbt] \begin{center}
 \resizebox{0.9\columnwidth}{!}{\includegraphics{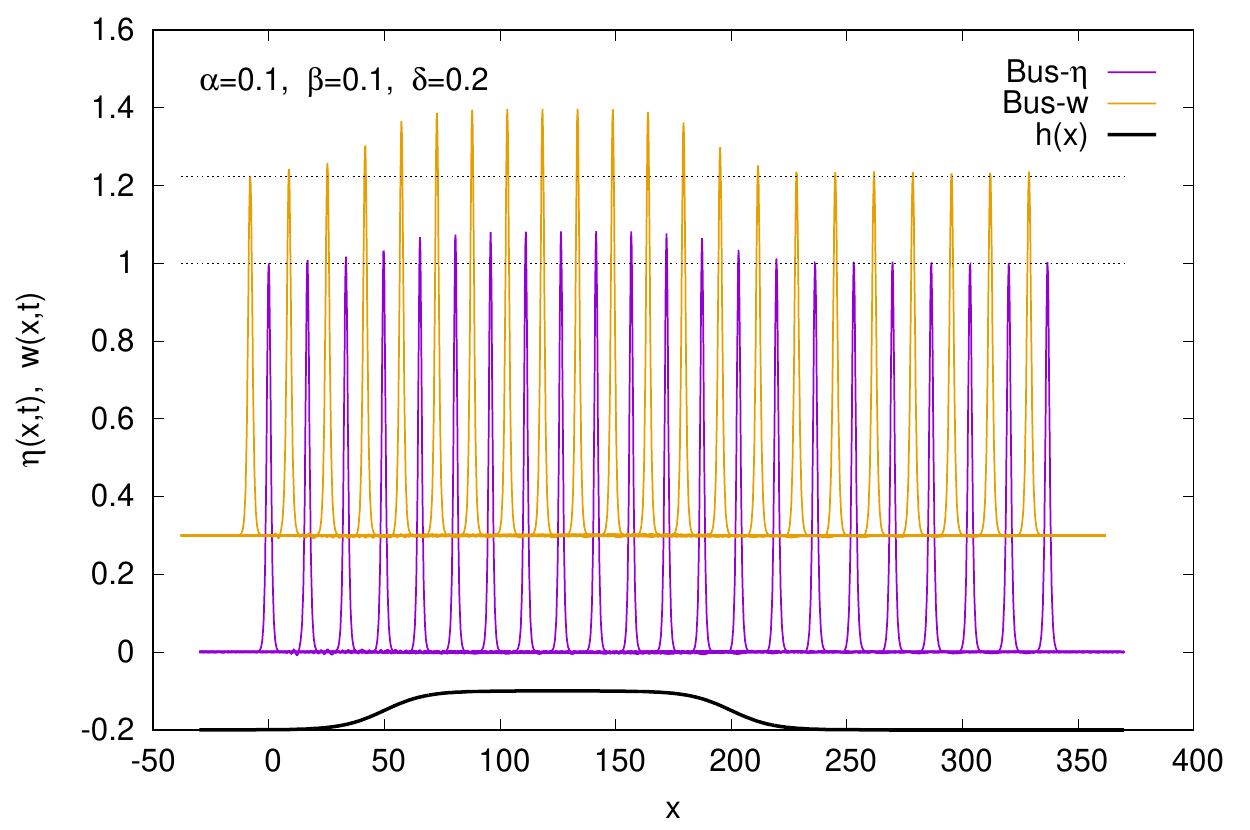}}
 \caption{Time evolution of the KdV soliton (\ref{1Sol})  and corresponding $w$ function obtained according to the Boussinesq set (\ref{4hx1})-(\ref{5hx1}) for the bottom given by (\ref{Btanh}) with $\delta=0.2$ and $\alpha=\beta=0.1$. Subsequent profiles correspond to times $t_n= n*16$, with $n=0,1,\ldots,21$. To avoid overlaps, profiles of $w$ function are shifted by 0.3 up and by 8 left.
 } \label{a1b1d2_BW.}
\end{center} \end{figure}

Next, we calculate the case in which the KdV soliton, formed on a flat bottom area enters the region over an extended bump of the shape given by the function 
\begin{equation} \label{Btanh}
h(x)=\frac{1}{2} \left(\textrm{\tnh}[0.055(x\!-\!50)]\!+\!\textrm{\tnh}[0.055(220\!-\!x)]\right).
\end{equation}

The results of numerical evolution of the KdV soliton according to the equation (\ref{kdvD}) (precisely, according to its discretized version (\ref{dkdv})) for the case $\alpha=\beta=0.1, \delta=0.2$ are presented in Fig.~\ref{a1b1d2KdV.}.  
Time separation between consecutive wave profiles is $dt=16$. These results show that according to the generalized KdV equation (\ref{kdvD}), the uneven bottom implies only minimal variations of solitons amplitude and velocity and creates a kind of small tail.

\begin{figure}[hbt] \begin{center}
 \resizebox{0.9\columnwidth}{!}{\includegraphics{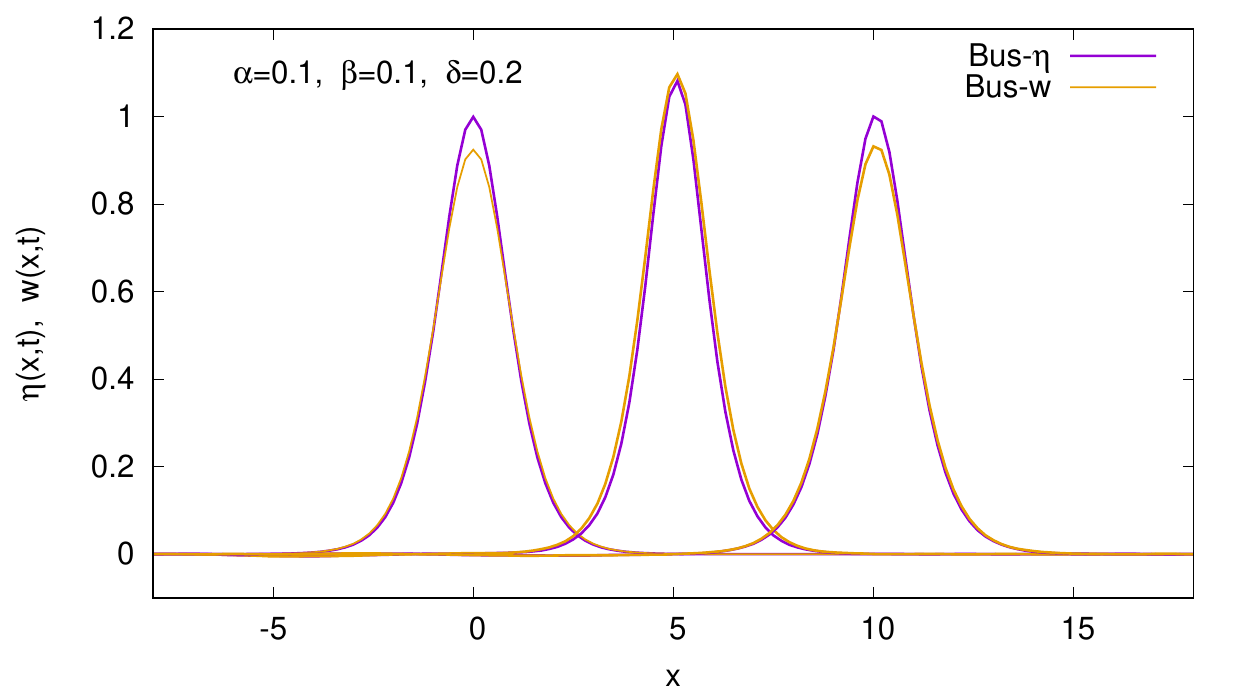}}
 \caption{Details of three profiles of $\eta(x,t)$ and $w(x,t)$ displayed in Fig.~\ref{a1b1d2_BW.} coresponding to time instants $t=0,128$ and 352. The second and third profile was shifted near the initial one for comparison. }
 \label{a1b1d2_BWr.}
\end{center} \end{figure}

In Fig.~\ref{a1b1d2_BW.}, we present the sequence of profiles obtained in numerics for the set of Boussinesq's equations (\ref{4hx1})-(\ref{5hx1}). Contrary to results from the KdV generalized for piecewise linear bottom function (\ref{kdvD}), in this case, we have almost ideal soliton shapes, without secondary soliton trains. Moreover, both $\eta$ and $w$ evolve similarly, with relative changes of $w$ bigger than those of $\eta$.
These relative changes are magnified in Fig.~\ref{a1b1d2_BWr.}. One has to stress that the changes in the surface wave amplitude and velocity obtained from the set of Boussinesq equations (\ref{4hx1})-(\ref{5hx1}) are substantially greater than those obtained from KdV equation (\ref{kdvD}), presented in Fig.~\ref{a1b1d2KdV.}. These properties of results remain similar for a wide range of parameters $\alpha,\beta$ when the bottom is the same. 
 
\section{Case $\alpha=O(\beta)$, ~$\delta=O(\beta^2)$ - generalization of KdV2
}  \label{2bus}

In this case (see details in \cite{KRcnsns}), from the boundary condition at the bottom we obtain 
\begin{equation} \label{Gd2}
F_x =  \beta\delta (h f_x)_x,
\end{equation}
valid up to fourth order in $\beta$
which inserted into (\ref{pot8}) gives the velocity potential valid up to fourth order
\begin{align} \label{pot16a}
 \phi= f &-\frac{1}{2}\beta z^2 f_{2x} + \frac{1}{24}\beta^2 z^4 f_{4x} - \frac{1}{720}\beta^3 z^6 f_{6x} \nonumber \\ &  
+  \beta^2\delta z (h f_x)_x +\frac{1}{40320}\beta^4 z^8 f_{8x} + O(\beta^5).
\end{align}
In principle, the Boussinesq equations can be consistently derived up to third order (remember term $-\frac{1}{\beta}\phi_x$ in (\ref{4BS})). However, we will proceed to second order, only.

Keeping only terms up to second order (for consistency with the order of approximation used in bottom boundary condition) one arrives at the second order Boussinesq set (see, \cite{KRcnsns}, eqs.~(37)-(38))
\begin{align} \label{4hx}
 \eta_t + w_x & +   \alpha(\eta w)_x-\frac{1}{6}\beta w_{3x}-\frac{1}{2} \alpha\beta (\eta w_{2x})_x  + \frac{1}{120}\beta^2 w_{5x} \nonumber \\  & -\delta(hw)_x  =  0, \\ \label{5hx}
w_t + \eta_x & + \alpha w w_x -\frac{1}{2}\beta\, w_{2xt} + \frac{1}{24}\beta^2\, w_{4xt}    \\  & 
+ \frac{1}{2} \alpha\beta\left(-2(\eta w_{xt})_x +  w_x w_{2x} - w w_{3x} \right)  =  0.  \nonumber
\end{align}

In the case of the flat bottom, that is when $\delta=0$, an appropriate form of $w$, precisely 
\begin{align} \label{wwabd}
w  = \eta & -\alpha\frac{1}{4}\eta^2 +\beta\frac{1}{3}\eta_{2x}+\alpha^2\frac{1}{8} \eta^3 \\ &
+ \alpha\beta \left( \frac{3}{16} \eta_{x}^2 + \frac{1}{2}\eta \eta_{2x}  \right)  +\beta^2 \frac{1}{10} \eta_{4x} \nonumber
 \end{align}
makes the equations (\ref{4hx})-(\ref{5hx}) identical. 
 The resulted equation is known as the {\it extended KdV} \cite{MS90} or 
{\it KdV2} \cite{KRbook} 
\begin{align} \label{nkdv2}	
\eta_t+\eta_x & +\alpha\frac{3}{2} \eta\eta_x+\beta\frac{1}{6}\eta_{3x} -\alpha^2\frac{3}{8}\eta^2\eta_x \\ & + \alpha\beta \left(\frac{23}{24}\eta_x\eta_{2x} +\frac{5}{12}\eta\eta_{3x} \right) + \beta^2\frac{19}{360}\eta_{5x}=0.  \nonumber
\end{align}
We proved recently that the extended KdV equations (\ref{nkdv2}), despite its nonintegrability, possesses three kinds of analytic solutions of the same form as the corresponding KdV solutions, with slightly different coefficients.
In \cite{KRI}, we found single soliton solution of the form $\eta(x,t)=A\,\textrm{\sech}[B(x-vt)]^2$. This form is the same as the form of the KdV soliton (\ref{1Sol}), but the coefficients are slightly different. In \cite{IKRR}, we found cnoidal solutions of the form $\eta(x,t)=A\,\textrm{\cn}[B(x-vt)]^2+D$ whereas in \cite{RKIsup,RKsup} we found so called 'superposition' periodic solutions of the form $\eta(x,t)=\frac{A}{2}(\textrm{\dn}^2[B(x-vt)]\pm\sqrt{m}\textrm{\cn}[B(x-vt)]\textrm{\dn}[B(x-vt)])$, where $\textrm{\cn},\textrm{\dn}$ are Jacobi elliptic functions.
It is worth to emphasize that contrary to the KdV case, exact multi-soliton solutions to the KdV2 do not exist \cite{KRappa19}.

\begin{figure}[htb] \begin{center}
 \resizebox{0.9\columnwidth}{!}{\includegraphics{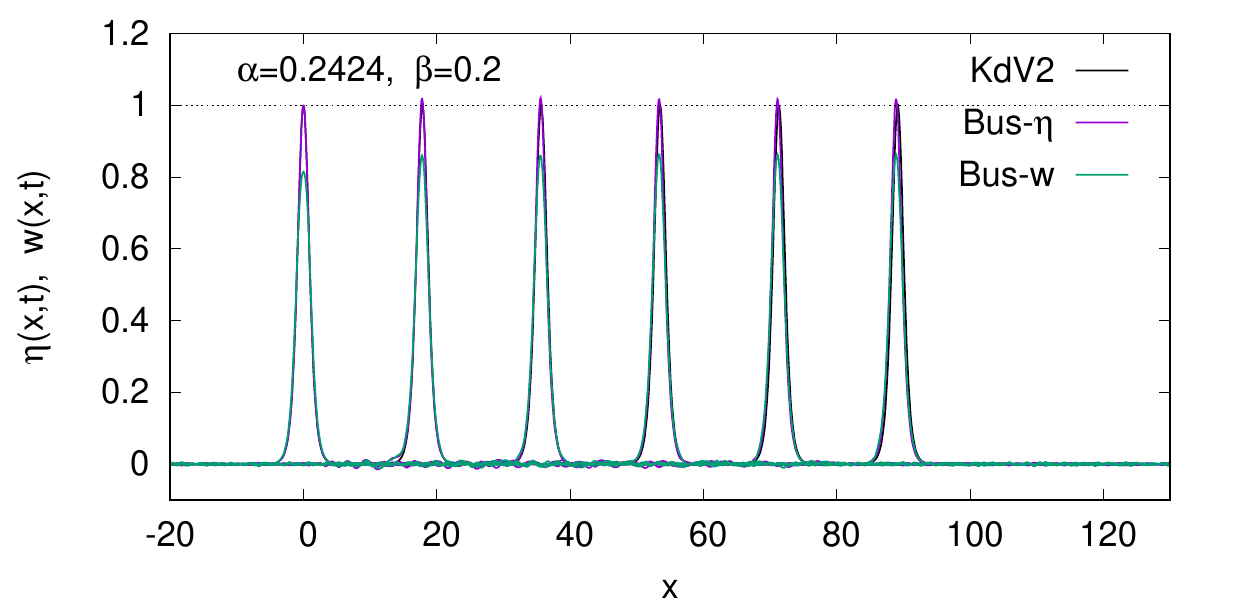}}
 \caption{The same as in Fig.\ \ref{a24b2d0K.}, but for the extended KdV (KdV2) equation (\ref{nkdv2}) and second order Boussinesq's set (\ref{4hx})-(\ref{5hx}) with $\delta=0$.} \label{a24b2BUS.}
\end{center} \end{figure}

Equations (\ref{4hx}) and (\ref{5hx}) can be made compatible only for when $h_{2x}=0$. In such case, the generalization of the KdV2 (\ref{nkdv2}) contains additional terms originating from the bottom variations (the bottom term is the same as in (\ref{kdvD}))
\begin{align} \label{nkdv2d}	
\eta_t+\eta_x & +\alpha\frac{3}{2} \eta\eta_x+\beta\frac{1}{6}\eta_{3x} -\alpha^2\frac{3}{8}\eta^2\eta_x\\ & +  \alpha\beta \left(\frac{23}{24}\eta_x\eta_{2x}  + \frac{5}{12}\eta\eta_{3x}\right)+ \beta^2\frac{19}{360}\eta_{5x}  \nonumber  \\ &  -  \frac{1}{4} \delta (2 h \eta_x+h_x \eta) =0. \nonumber
\end{align}

In numerical calculations, we use the same FDM method as that described by equations (\ref{dkdv})-(\ref{dw}), extended by including appropriate terms, second order in small parameters. 
As initial condition for $\eta(t=0)$ the KdV2 solitons are used, whereas the initial condition for $w$ is given by (\ref{wwabd}) with substitution $\eta=\eta(t=0)$.
So, for the evolution shown in Fig.~\ref{a24b2BUS.} the initial condition has the the same form (\ref{1Sol})  but with coefficients: $A\approx \frac{0.2424}{\alpha},~B\approx\sqrt{0.6\,\frac{\alpha}{\beta}A}$ and $v\approx 1.11455$. The parameter $\alpha=0.2424$ assures the amplitude equal one.

\begin{figure}[hbt] \begin{center} \resizebox{0.9\columnwidth}{!}{\includegraphics{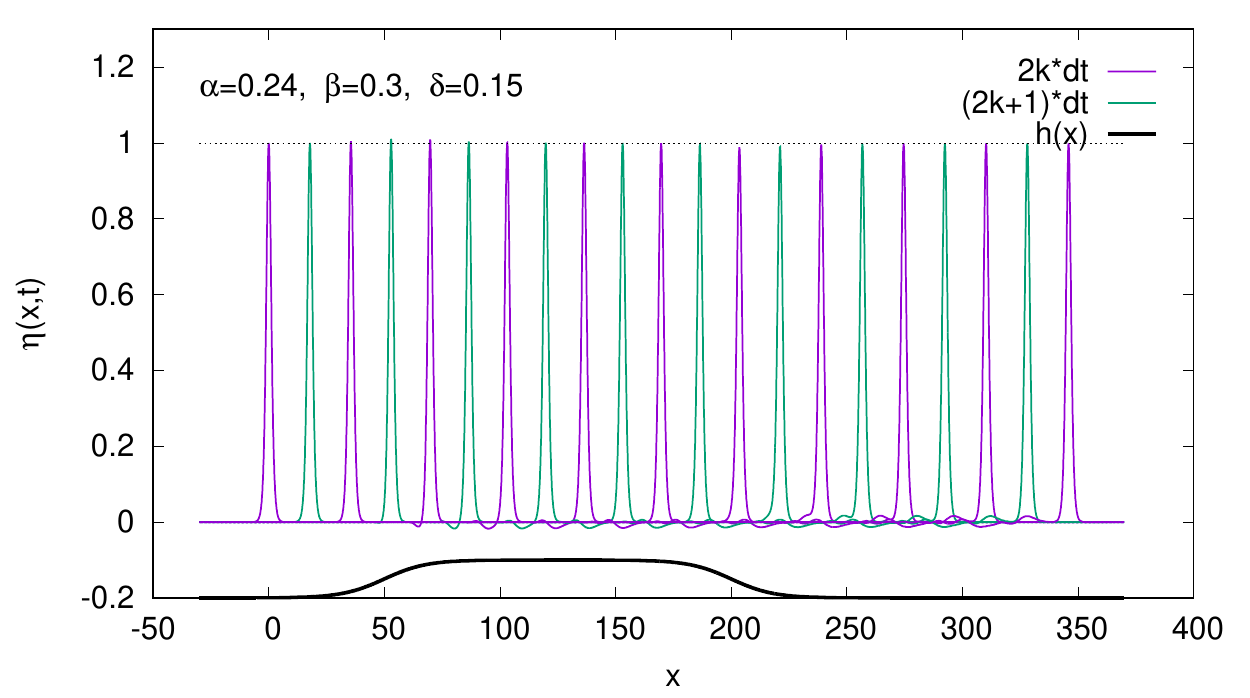}} \caption{The same as in Fig.\ \ref{a1b1d2KdV.}, but for the extended KdV (KdV2) equation (\ref{nkdv2d}) and $\delta=0.15$.
 } \label{a24b3d15K.}
\end{center} \end{figure}

Now, we will compare the time evolution of the KdV2 soliton, obtained according to second order equations (KdV2 or {\it extended KdV}).
In Fig.~\ref{a24b3d15K.}, we display profiles of KdV2 soliton, which enters the region of the uneven bottom. The time evolution is obtained from the generalized KdV2 equation (\ref{nkdv2d}). The behavior of solutions, despite different values of small parameters, remains very similar to that presented in Fig.~\ref{a1b1d2KdV.} for the first order equation.

\begin{figure}[htb] \begin{center} \resizebox{0.9\columnwidth}{!}{\includegraphics{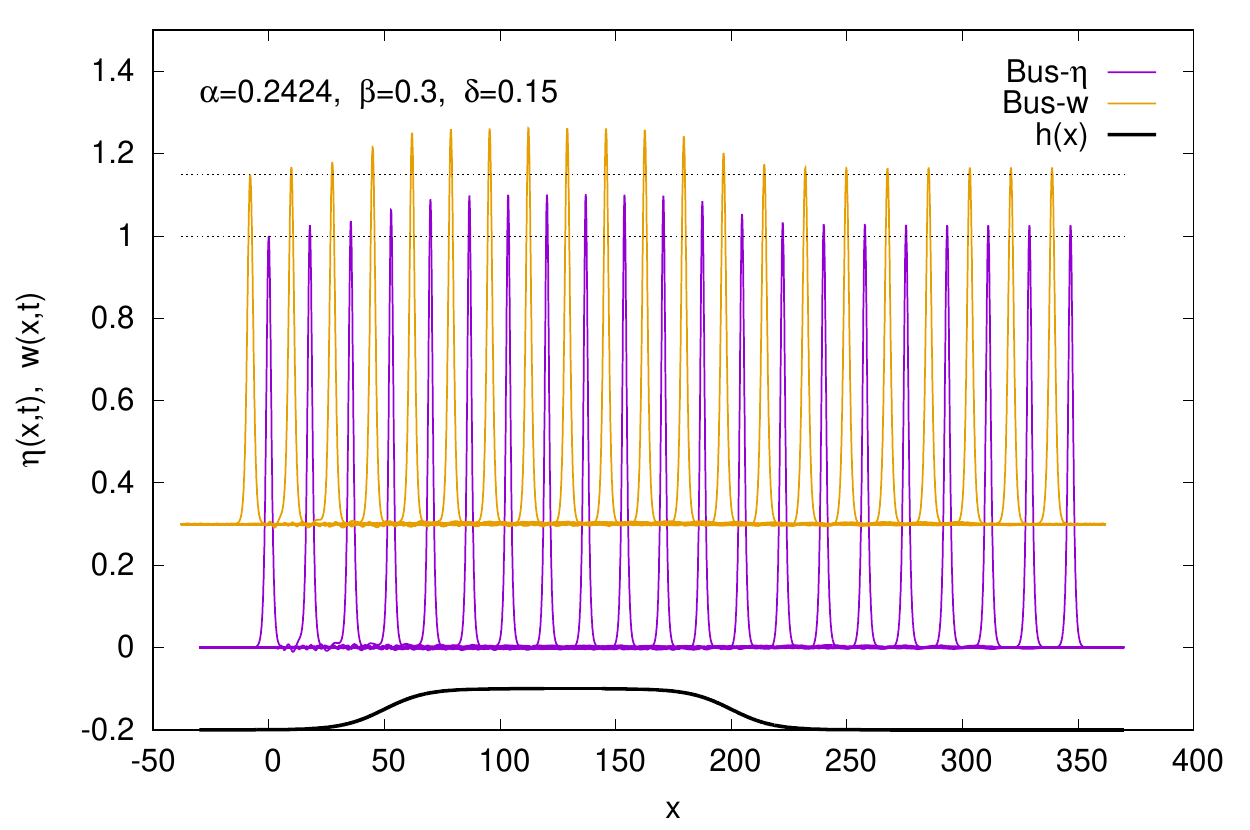}} \caption{ The same as in Fig.\ \ref{a1b1d2_BW.}, but for the  second order Boussinesq's set (\ref{4hx})-(\ref{5hx}). } \label{a24b3d15BW.}
\end{center} \end{figure}

In Fig.~\ref{a24b3d15BW.}, the initial KdV2 soliton evolves according to second order Boussinesq's equations (\ref{4hx})-(\ref{5hx}). In this case, similarly as in Fig.~\ref{a1b1d2_BW.}, one observes the much greater influence of the bottom variation on changes of soliton's amplitude and velocity.

\section{Case $\alpha=O(\beta^2)$, ~$\delta=O(\beta^2)$ - generalization of fifth-order KdV equation}  \label{5bus}

In this case, since $\delta =O(\beta^2)$, the forms of the function $F_x$ and the velocity potential are given by (\ref{Gd2})-(\ref{pot16a}). 
Keeping only terms up to second order  one arrives at the second order Boussinesq system (see, \cite{KRcnsns}, eqs.\ (61)-(62))
\begin{align} \label{3Ba2d2}
\eta_t+w_x &-\frac{1}{6}\beta\, w_{3x} +\alpha (w\eta)_x +\frac{1}{120}\beta^2 w_{5x}-\delta (h w)_x=0, 
\\ \label{4Ba2d2}
w_t+\eta_x & -\beta\left(\frac{1}{2} w_{2xt}+\tau \eta_{3x}\right)
+ \alpha\, w w_x+ \frac{1}{24}\beta^2w_{4xt}= 0. 
\end{align}
Here, one has to keep terms from surface tension $\tau\ne 0$. These terms are important because for the flat bottom ($\delta=0$), the equations (\ref{3Ba2d2})-(\ref{4Ba2d2}) can be made compatible leading to so-called \emph{fifth-order KdV equation}  derived by Hunter and Sheurle in \cite{HS88} as a model equation for gravity-capillary shallow water waves of small amplitude. 

Similarly like in the previous sections for uneven bottom, the equations (\ref{3Ba2d2})-(\ref{4Ba2d2}) can be made compatible only when the bottom function is piecewise linear. The resulting wave equation, a generalization of the fifth-order KdV equation has the following form (see, eq.\ (68) in \cite{KRcnsns})
\begin{align}\label{5kdvQ}
\eta_t+\eta_x  & + \frac{3}{2}\alpha\eta\eta_x +\beta \frac{1-3\tau}{6} \eta_{3x} 
 +\beta^2 \frac{19-30\tau-45\tau^2}{360}\eta_{5x} \nonumber  \\ &
-\frac{1}{4} \delta (2 h \eta_x+h_x \eta) =0.
\end{align}
The equation (\ref{5kdvQ}) differs from the fifth-order KdV equation by the last term only.

In numerical simulations, we again want to compare the time evolution of surface waves obtained from the single wave equation (\ref{5kdvQ}) with time evolution obtained from the Boussinesq set (\ref{3Ba2d2})-(\ref{4Ba2d2}).

It is well known, see, e.g.~\cite{Dey96,Bri02}, that the fifth order KdV equation has a  soliton solution in the form
\begin{equation} \label{5sol1}
\eta(x,t) = A\, \text{\sech}^4[B(x-vt)].
\end{equation}
For the fifth order KdV equation in the form (\ref{5kdvQ}) one obtains the following values of the coefficients:\\
\begin{align} \label{5sol1AB}
A & =\frac{700 (1 - 3 \tau)^2}{169 (-19 + 30 \tau + 45 \tau^2) \alpha},    \\
B & =\sqrt{ \frac{15 (1 - 3 \tau)}{13 (-19 + 30 \tau + 45 \tau^2) \beta}} \nonumber 
\end{align}
and 
\begin{equation} \label{5sol1V}
v= \frac{-2851 + 2910 \tau + 10845 \tau^2}{169 (-19 + 30 \tau + 45 \tau^2)}.
\end{equation}

Real solutions require $\tau>\frac{1}{3}$. 
Using $\tau=0.35$ we obtain ~$A\approx -0.00346612/\alpha$, ~$B\approx 0.0193112/\beta$~ and ~$v\approx 0.998217$.
To begin evolution according to the Boussinesq equations one needs the initial condition for $w$ function which has the following form
\begin{equation} \label{5sol1w}
w(x,t) = \eta + \beta \frac{2-3\tau}{6} \eta_{2x} - \frac{1}{4} \alpha \eta^2 +\beta^2 \frac{12-20\tau-15\tau^2}{120} \eta_{4x}.
\end{equation}

\begin{figure}[htb] \begin{center}
\resizebox{0.9\columnwidth}{!}{\includegraphics[angle=270]{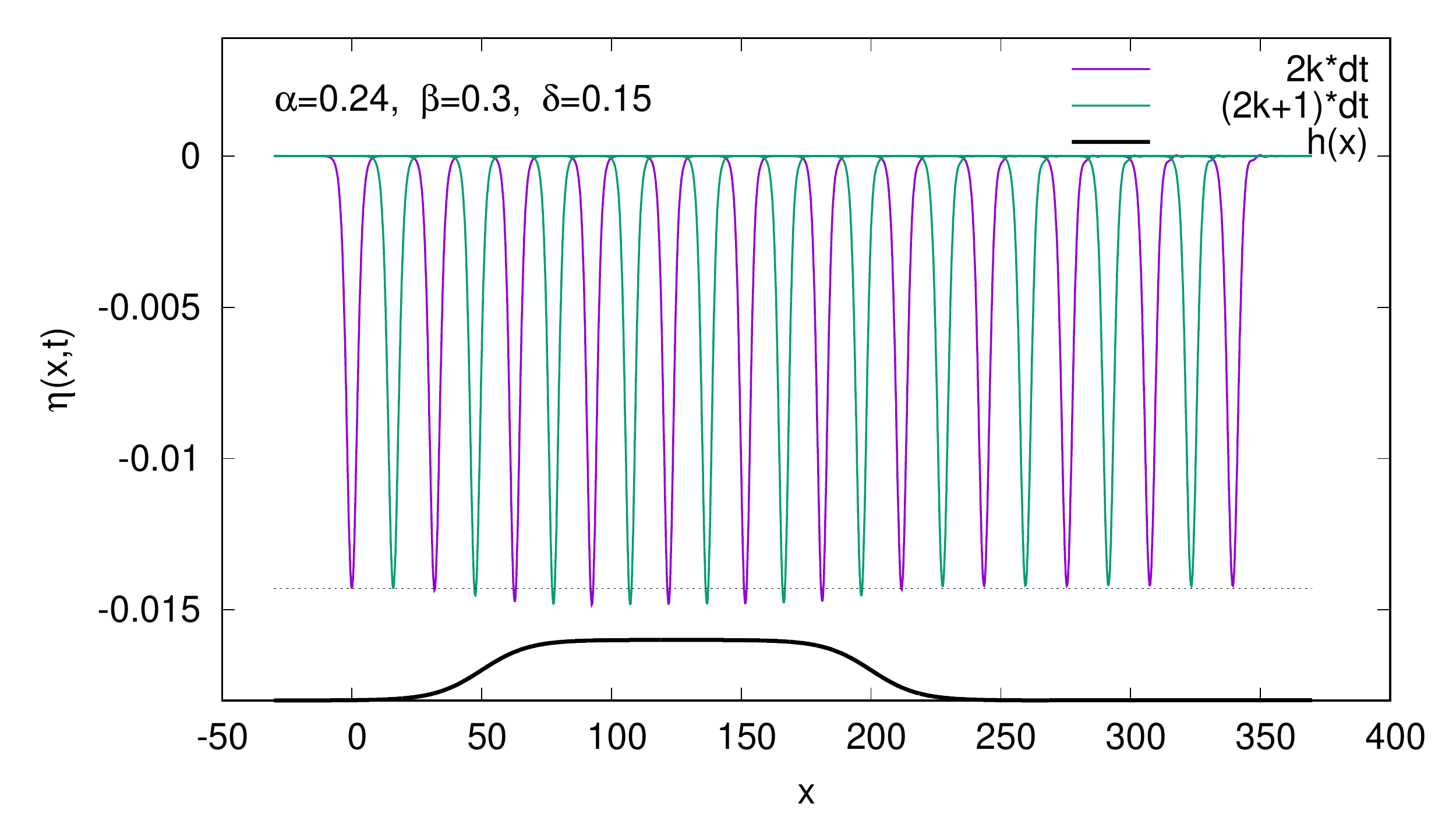}} \caption{ The same as in Fig.~\ref{a1b1d2KdV.} but for the 5th-order KdV equation (\ref{5kdvQ}). } \label{F232.}\end{center} \end{figure}

\begin{figure}[htb] \begin{center} \resizebox{0.9\columnwidth}{!}{\includegraphics[angle=270]{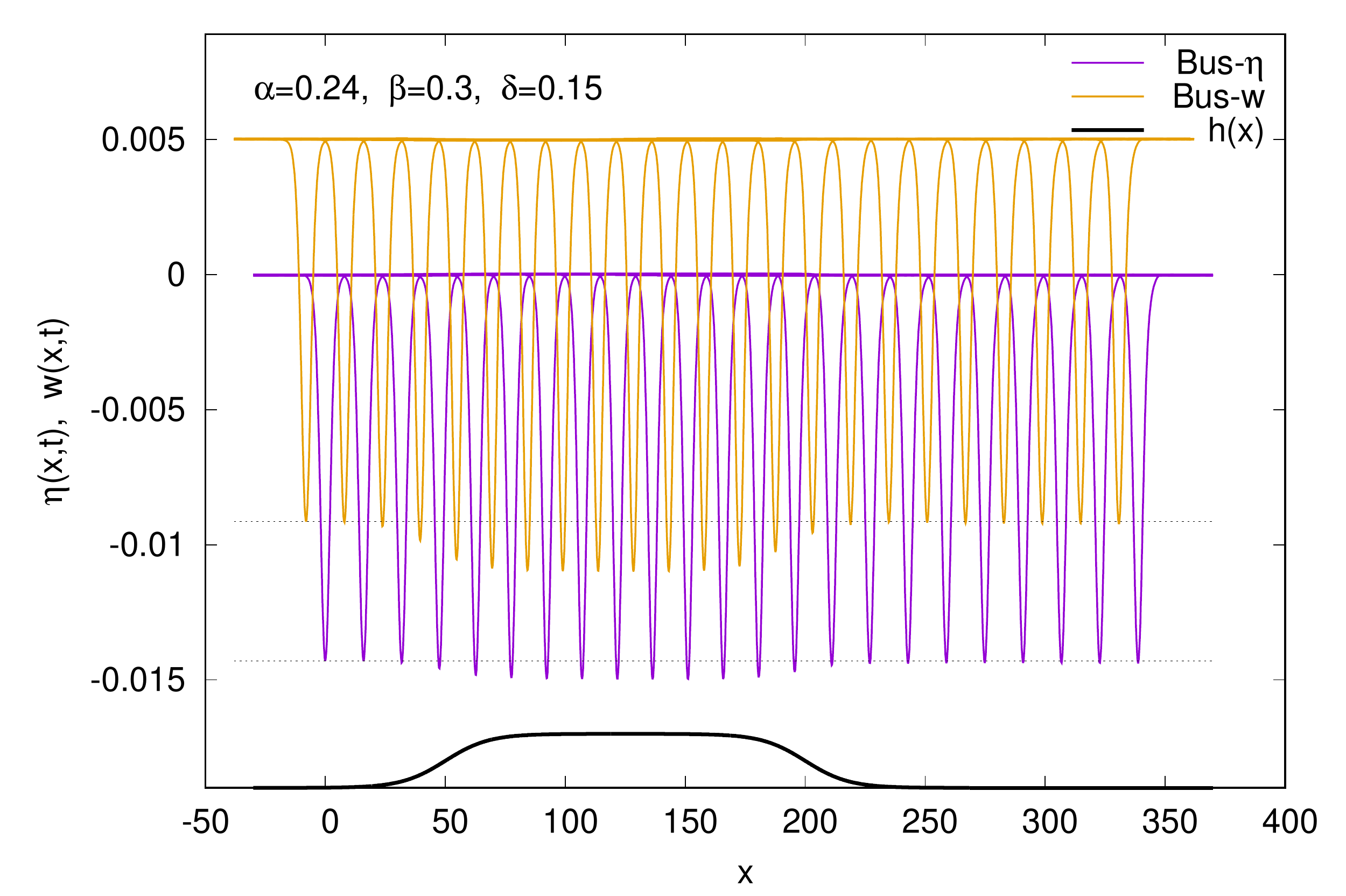}} \caption{Profiles of functions $\eta$ and $w$ obtained from the second order Boussinesq  set (\ref{3Ba2d2})-(\ref{4Ba2d2}). In order to avoid overlaps profiles of $w$ function are shifted by 0.005 up and by 8 left.
 } \label{F230-1.} \end{center} \end{figure}

The numerical results of the time evolution of 5th-order KdV soliton according to equation (\ref{5kdvQ}) are presented in Fig.~\ref{F232.}.
The evolution of the same initial 5th-order KdV soliton according to Boussinesq's equations (\ref{3Ba2d2})-(\ref{4Ba2d2}) is displayed in Fig.~\ref{F230-1.}.
Similarly, as in the previous section, the impact of the bottom variation on the surface wave manifests more evident in the case of Boussinesq's equations.

\section{Case $\beta=O(\alpha^2)$, ~$\delta=O(\alpha^2)$ - generalization of the Gardner equation 
}  \label{GE}

In this case, the leading parameter is parameter $\alpha$. The boundary condition at the bottom requires 
\begin{align*}
F_x  -  \delta (hf_x)_x & + \frac{1}{2} \beta\delta^2 (h^2 F_{2x})_x 
+ O( \alpha^{8}) =0.
\end{align*}
Neglecting higher order terms we can use 
\begin{equation} \label{Fx2}
F_x = \delta  (hf_x)_x +O( \alpha^{6}),
\end{equation}
which ensures the expression of $\phi$ through only one unknown function $f$ and its derivatives. 
Now, the Boussinesq set (up to second order) is given by (see, eqs.\ (85)-(86) in \cite{KRcnsns})
\begin{align} \label{bbu3}
\eta_t+ w_x  &  + \alpha (\eta w)_x - \frac{1}{6} \beta\, w_{3x} - \delta (h w)_x  
=0, \\  \label{bbu4}
w_t + \eta_x & + \alpha w w_x -\beta \left(\tau \eta_{3x}+  \frac{1}{2} w_{2xt} \right) 
=0. 
\end{align}

Formally, the equations (\ref{bbu3})-(\ref{bbu4}) are identical to the equations (\ref{4hx1})-(\ref{5hx1}) obtained for the case $ \alpha \approx \beta \approx \delta $, that is 1st order equations that lead to the KdV equation when $ \delta = 0 $. This suggests that the solutions  $ \eta, w $ of the system of equations (\ref{bbu3})-(\ref{bbu4}) may have identical functional form to those from the equation KdV.

Similarly, as in the previous sections for the uneven bottom, the equations (\ref{bbu3})-(\ref{bbu4}) can be made compatible only when the bottom function is piecewise linear. The resulting wave equation, a generalization of the Gardner equation has the following form (see, eq.\ (91) in \cite{KRcnsns})
\begin{align} \label{GardH}
\eta_t+\eta_x  +\frac{3}{2} \alpha  \eta\eta_x & +\alpha^2 \left(- \frac{3}{8} \eta^2\eta_x\right) + \frac{1-3\tau}{6} \beta\, \eta_{3x}  \nonumber \\ & 
-\frac{1}{4}\delta (2 h \eta_x+h_x \eta) =0 .
\end{align}
Setting $\delta=0$ gives the well known Gardner equation (combined KdV-mKdV equation)
\begin{align} \label{Gard}
\eta_t+\eta_x +\frac{3}{2} \alpha  \eta\eta_x & +\alpha^2 \left(- \frac{3}{8} \eta^2\eta_x\right) + \frac{1-3\tau}{6} \beta\, \eta_{3x}  =0 .
\end{align}
In this case the $w$ function, limited to second order terms is, (see, e.g.~\cite[Eq.~(A.1)]{BurSerg})
\begin{align} \label{GardW}
w = \eta-\frac{1}{4}\alpha\eta^2+\frac{1}{8}\alpha^2\eta^3+\frac{2-3\tau}{6}\beta \eta_{2x}.
\end{align}

It is well known, e.g.~\cite{GPT99,OPSS15}, that for the Gardner equation (\ref{Gard}) there exists one parameter family of analytic solutions in the form
\begin{equation} \label{1sOPSS}
\eta(x,t)= \frac{A}{1+ B\, \textrm{cosh}[(x-v\,t)/\Delta]}.
\end{equation}
The equation (\ref{Gard}) imposes three conditions on coefficients $A,B,v,\Delta$ of solutions. So, three of them can be expressed as functions of the single one. Choosing $\Delta$ as the independent parameter one obtains the following relations 
\begin{equation} \label{Gpar}
A=\frac{2\beta}{3\alpha}\frac{1}{\Delta^2}, \quad B= \pm\sqrt{1-\frac{\beta}{6}\frac{1}{\Delta^2}}, \quad V=1+\frac{\beta}{6}\frac{1}{\Delta^2}.
\end{equation} 

Soliton's amplitude is then $$\displaystyle \eta_0=\frac{A}{1+B}=\frac{2\beta}{3\alpha\,\Delta^2\left(1\pm \sqrt{1-\frac{\beta}{6}\frac{1}{\Delta^2}}\right)}.$$
For $B\in \mathbb{R}$,~ ~$\Delta^2\ge \frac{\beta}{6}$. Assuming $B\ge 0$ one has limiting values of $B$ as $B=0$, when $\Delta^2=\frac{\beta}{6}$, and $B=1$, when $\Delta^2\to\infty$. So, the corresponding limiting values of the amplitude are $\eta_0=\frac{4}{\alpha}$ and $\eta_0=0$, respectively.
The equations (\ref{Gpar}) are obtained by setting $\tau=0$ in (\ref{Gard}), which is a fair approximation for surface gravity waves. 

\subsection{Gardner equations for shallow water waves} \label{GBlt1}

\begin{figure}[htb] \begin{center} \resizebox{0.9\columnwidth}{!}{\includegraphics[angle=270]{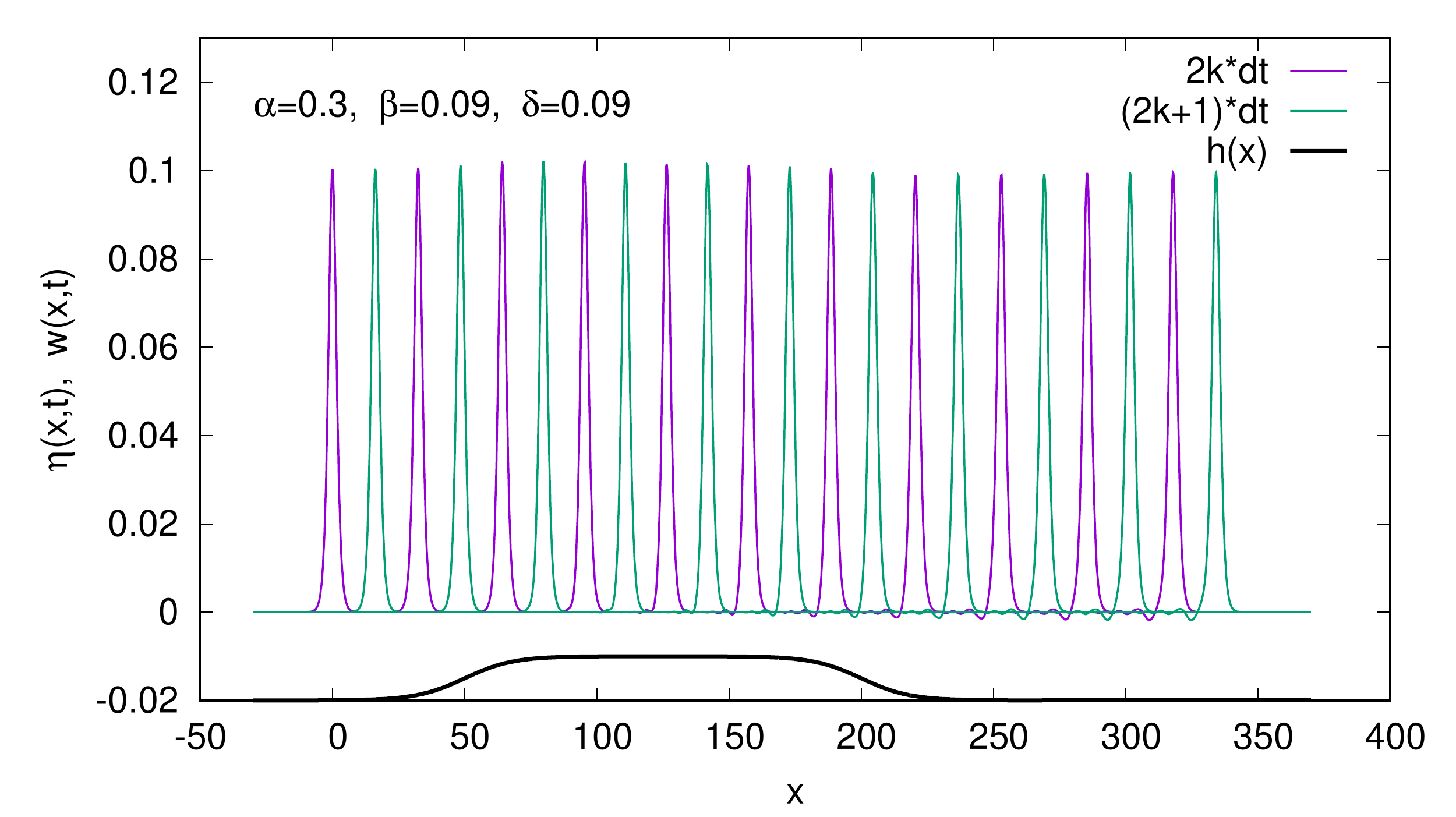}} \caption{ The same as in Fig.~\ref{a1b1d2KdV.} but for the Gardner equation (\ref{GardH}).  Parameters $\alpha=0.3, \beta=\delta=0.09, \tau=0$ of the equation were used. The value $\Delta^2=1$ was chosen for the initial soliton (\ref{1sOPSS}).}\label{a3bd09G.}\end{center} \end{figure}

\begin{figure}[bht] \begin{center} \resizebox{0.9\columnwidth}{!}{\includegraphics[angle=270]{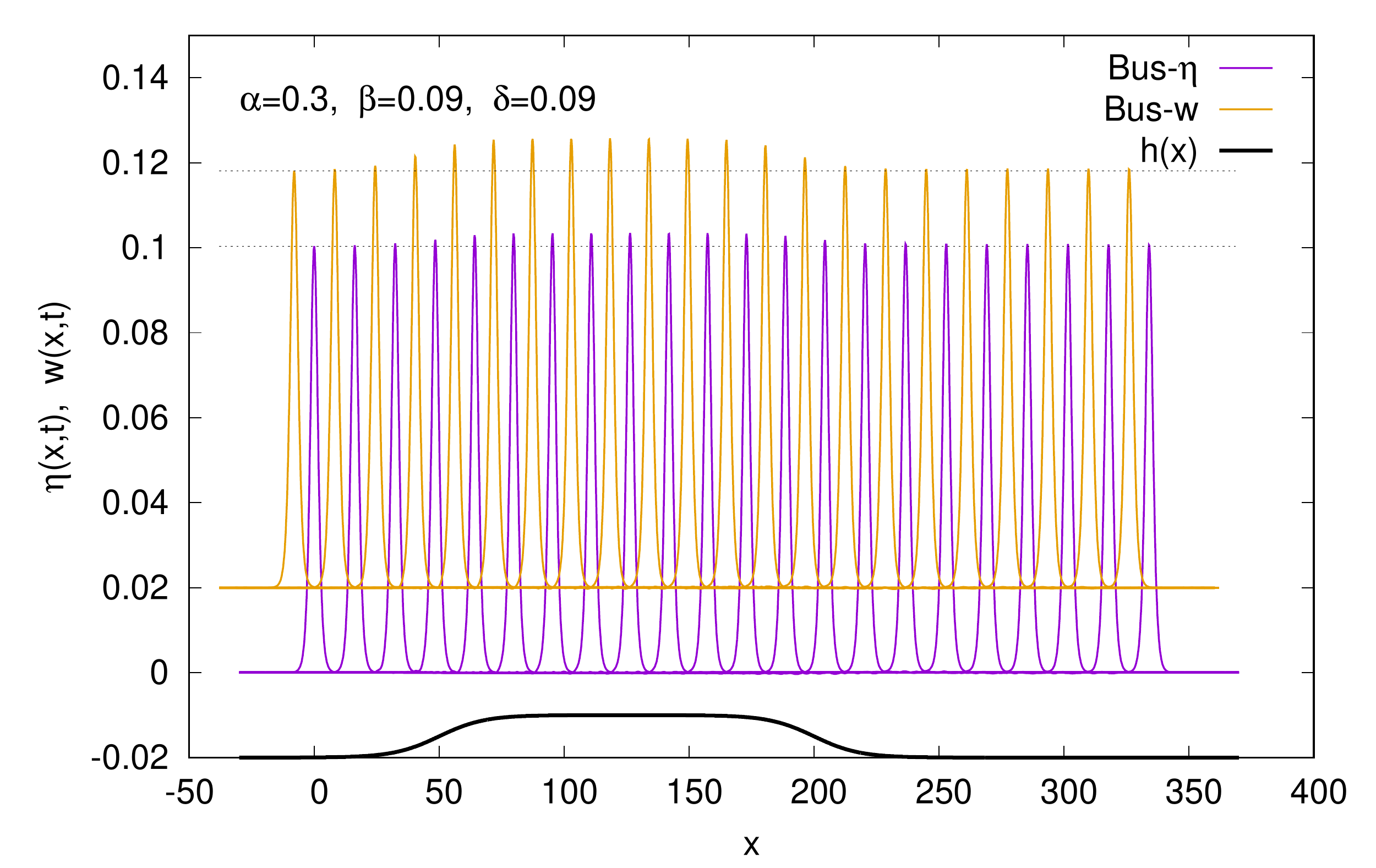}} \caption{Profiles of functions $\eta$ and $w$ obtained from the second order Boussinesq's set (\ref{bbu3})-(\ref{bbu4}), the precursors of the Gardner equation. Parameters are the same as in Fig.~\ref{a3bd09G.}. In order to avoid overlaps, profiles of $w$ function are shifted by 0.005 up and by 8 left. } \label{a3bd09BW.}\end{center} \end{figure}

Let us recall, that the Gardner equation (\ref{GardH}) and (\ref{Gard}) have been derived under assumptions that parameter $\alpha$ is small and parameters $\beta$ and $\delta$ are of one order smaller, that is $\beta\approx \delta \approx O(\alpha^2)$. Therefore, for numerical simulations we take ~$\alpha=0.3$, ~$\beta=0.09$, ~$\delta=0.09$. These values of $\alpha,\beta$ imply  ~$A= 0.2/\Delta^2$,
~$B=\sqrt{1-\frac{0.015}{\Delta^2}} $,~ and ~$V= 1+\frac{0.015}{\Delta^2}$. In Fig.~\ref{a3bd09G.} we display profiles of Gardner's soliton obtained during the motion according to the Gardner equation (\ref{Gard}). These results can be compared with the evolution of the same initial Gardner's soliton according to the Boussinesq equations (\ref{bbu3})-(\ref{bbu4}), shown in Fig.~\ref{a3bd09BW.}. In the last case the initial condition for the $w$ function is taken in the form (\ref{GardW}).

\subsection{Gardner equation for thin liquid layers}\label{GBgt1}

\begin{figure}[htb] \begin{center}
\resizebox{0.9\columnwidth}{!}{\includegraphics[angle=270]{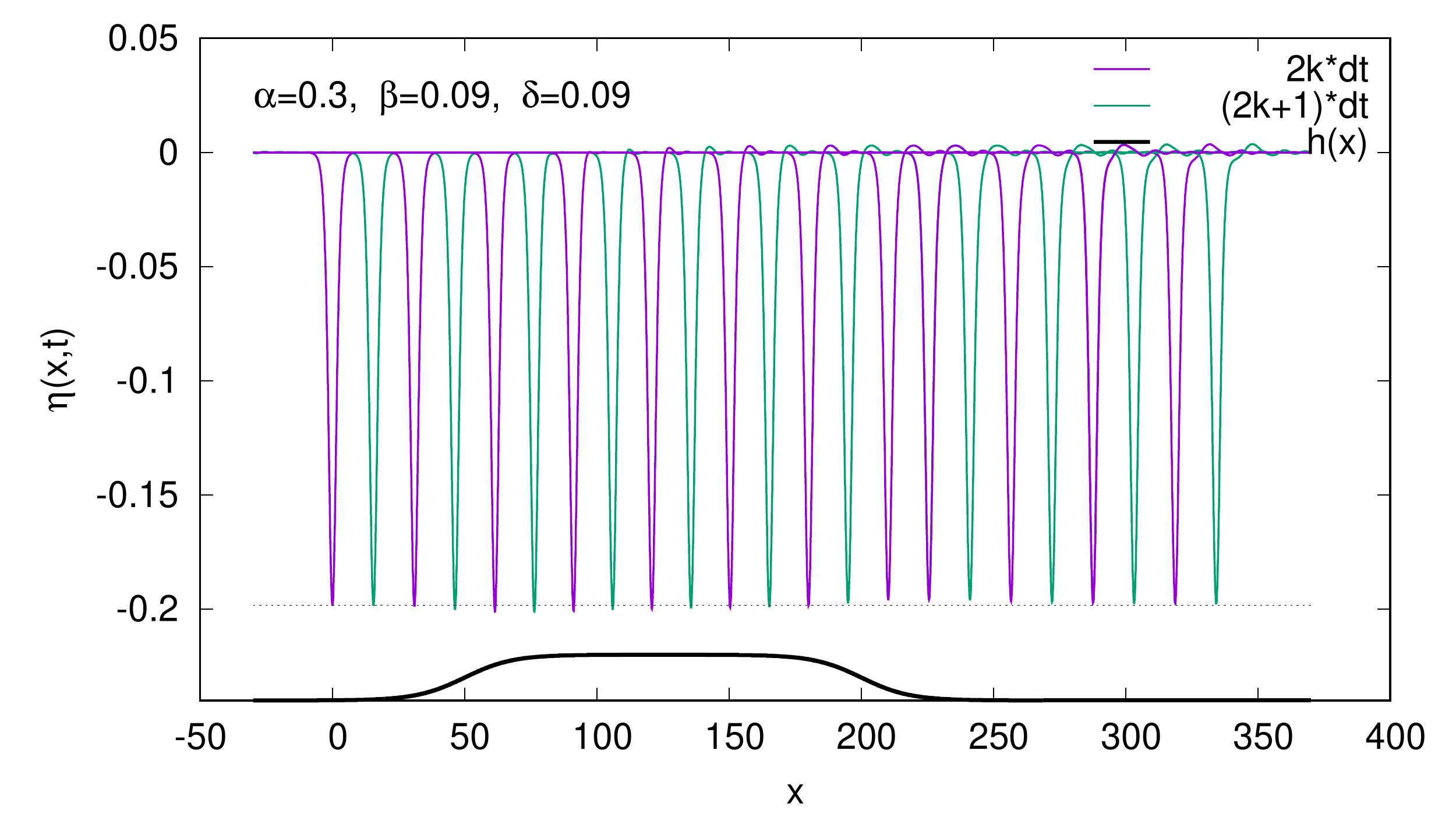}} \caption{ The same as in Fig.~\ref{a1b1d2KdV.} but for the Gardner equation (\ref{GardH}).  Parameters $\alpha=0.3, \beta=\delta=0.09, \tau=1$ of the equation were used. The value $\Delta^2=1$ was chosen for the initial soliton (\ref{1sOPSS}).}\label{a3bd09Gt.}\end{center} \end{figure}

\begin{figure}[bht] \begin{center} \resizebox{0.9\columnwidth}{!}{\includegraphics[angle=270]{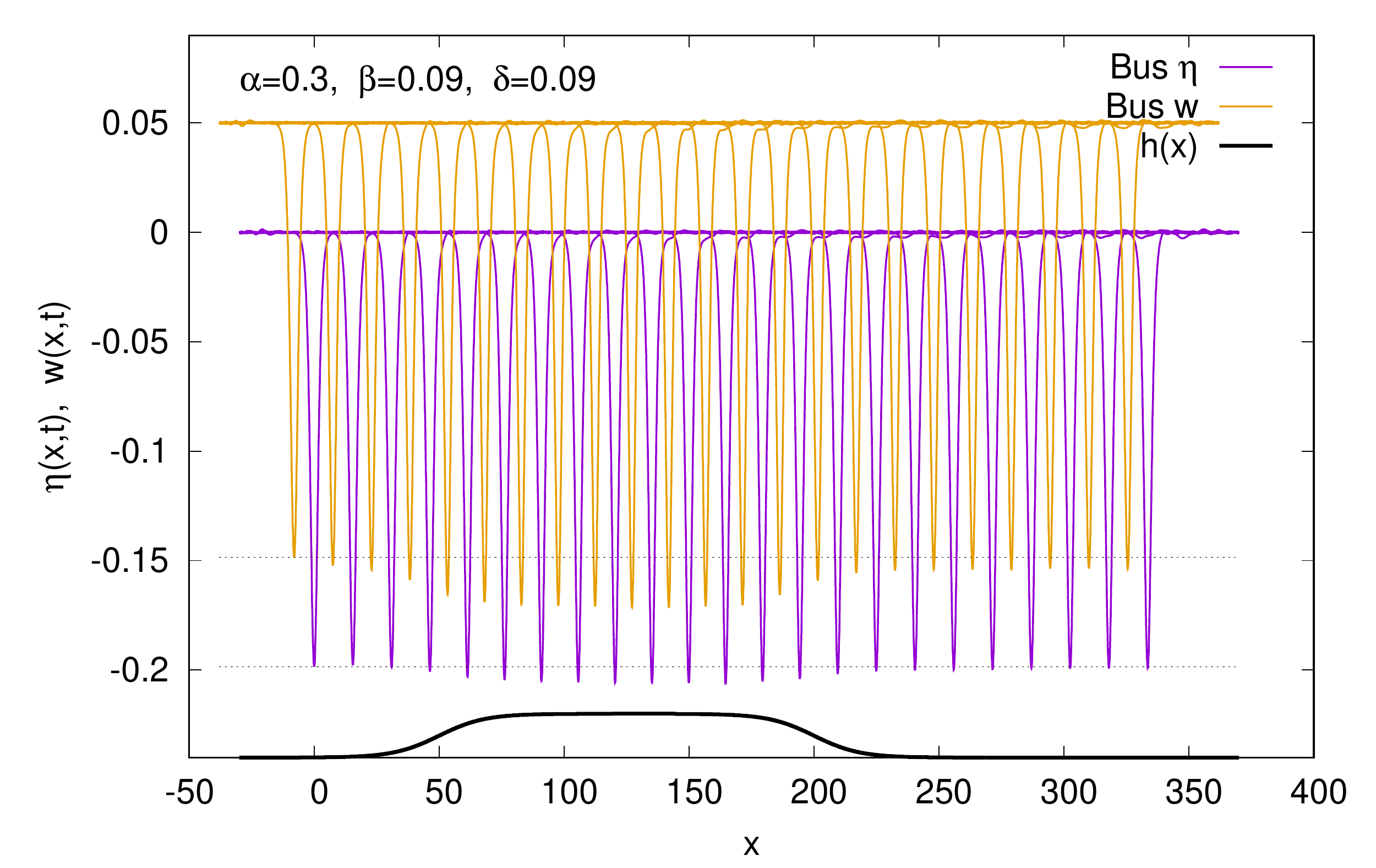}} \caption{Profiles of functions $\eta$ and $w$ obtained from the second order Boussinesq's set (\ref{bbu3})-(\ref{bbu4}), the precoursors of the Gardner equation. Parameters are the same as in Fig.~\ref{a3bd09G.}. In order to avoid overlaps, profiles of $w$ function are shifted by 0.005 up and by 8 left. } \label{a3bd09GBt.}\end{center} \end{figure}

In this case we have to take into account that the Bond number $\tau$ can be greater than 1/3. Then the coefficient $\frac{1-3\tau}{6}\beta$ in eq.~(\ref{Gard}) can become negative and the parameter $B$ can be greater that 1. The parameters of the solution (\ref{1sOPSS}) are now
\begin{align} \label{GparT}
A & =\frac{2(1-3\tau)\beta}{3\alpha\,\Delta^2}, \qquad B= \pm\sqrt{1-\frac{(1-3\tau)\beta}{6\,\Delta^2}}, \nonumber\\   V & =1+\frac{(1-3\tau)\beta}{6\,\Delta^2},
\end{align} 
with soliton's amplitude given by
$$\eta_0=\frac{A}{1+B}=\frac{2(1-3\tau)\beta}{3\alpha\,\Delta^2\left(1\pm \sqrt{1-\frac{(1-3\tau)\beta}{6\,\Delta^2}}\right)}.
$$
The examples of time evolution of Gardner's soliton for the uneven bottom are displayed in Figs.~\ref{a3bd09Gt.} and \ref{a3bd09GBt.}. In Fig.~\ref{a3bd09Gt.} we present results obtained from the Gardner equation (\ref{Gard}), whereas in Fig.~\ref{a3bd09GBt.} those which result from Boussinesq's set (\ref{bbu3})-(\ref{bbu4}). The time step between subsequent profiles is 16. In both cases we used the same initial condition in the form of Gardner's soliton (\ref{1sOPSS}) with parameters $A,B,V$ given by (\ref{GparT}). For the Boussinesq system (\ref{bbu3})-(\ref{bbu4}) the initial condition for the $w$ function is taken in the form (\ref{GardW}).

Comparing Figs.~\ref{a3bd09G.}-\ref{a3bd09GBt.} we recognize the same qualitative properties as in previous sections. The impact of bottom changes on surface waves is more prominent when the evolution proceeds according to the Boussinesq equations than in the case of the single Gardner equation.

\section{Non-soliton initial conditions}\label{nonSol}

In all examples presented in previous sections, the initial conditions were chosen in the form of soliton solutions to particular wave equations. Such initial conditions appear to be extremely resistant to disturbances introduced by varying bottom. This means that a bottom with a small amplitude introduces only small changes of soliton’s amplitude and velocity, leaving the shape almost unchanged. On the other hand, in all considered cases, the impact of the bottom variations on the changes of surface waves is distinctly more significant when calculated from the Boussinesq equations than when calculated from single wave equations. 

Now, we study some examples of the time evolution of initial waves (elevation or depression), which shapes are different from solitons of particular equations.
We study these evolutions taking the initial shape of the wave in the form of a Gaussian with the amplitude equal to soliton’s amplitude but with the width providing the volume of the deformation being substantially greater than that of a soliton. In particular, we focus on the case, which, for the flat bottom, leads to the extended KdV equation (KdV2). In all other cases, the behavior of the evolution of wave profiles appears qualitatively to be very similar.

\subsection{KdV case} \label{kdv-inv}

\begin{figure}[bht] \begin{center}
\resizebox{0.9\columnwidth}{!}{\includegraphics{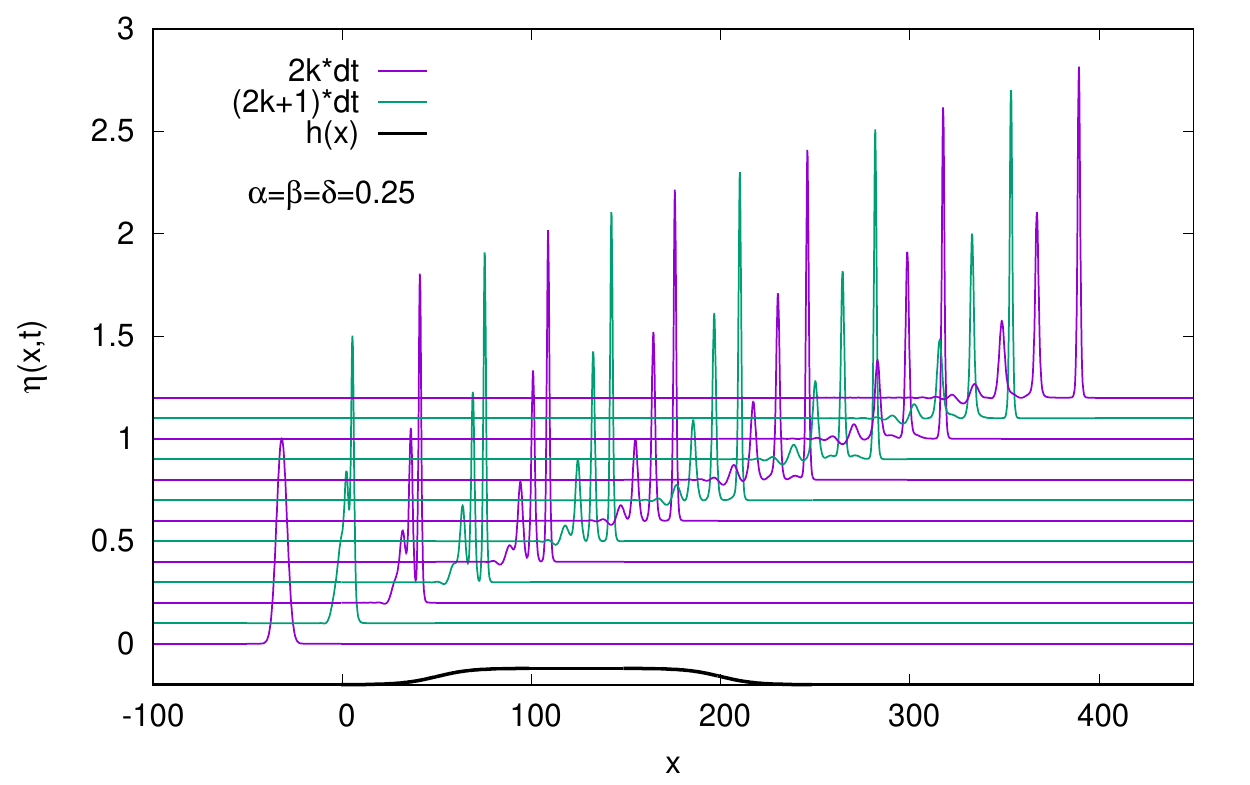}}
 \caption{Time evolution obtained according to the KdV equation 
(\ref{kdvD}).~Initial Gaussian profile with the triple volume of the KdV soliton, the same velocity and amplitude.  Here, time step between the consecutive profiles is $dt=32$.} \label{abd15_Ks3.}\end{center} \end{figure}

\begin{figure}[bht] \begin{center} \resizebox{0.9\columnwidth}{!}{\includegraphics{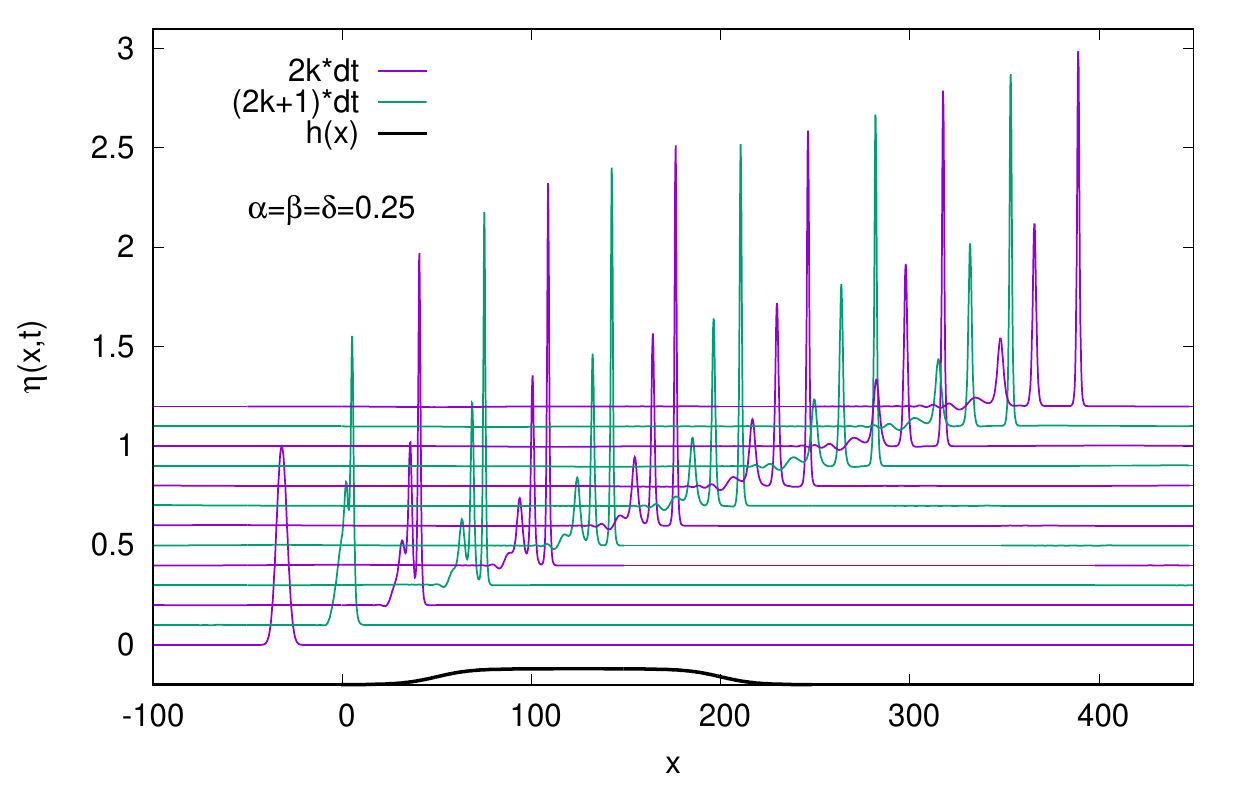}}
 \caption{Time evolution obtained according to Boussinesq's equations (\ref{4hx1})-(\ref{5hx1}).
Initial Gaussian profile with the triple volume of the KdV soliton, the same velocity and amplitude.  Here, time step between the consecutive profiles is $dt=32$. } \label{abd15_Bs3.}\end{center} \end{figure}

In Figs.~\ref{abd15_Ks3.} and \ref{abd15_Bs3.} we show the profiles of the time evolution of waves calculated according to equations (\ref{kdvD}) (KdV generalized for an uneven bottom) and (\ref{4hx1})-(\ref{5hx1}) (the corresponding Boussinesq equations), respectively. In both cases, the initial condition was taken as the Gaussian profile moving with the KdV soliton's velocity, the same amplitude, but with the triple volume of the fluid distortion from equilibrium. The parameters of wave equations are $\alpha=\beta=\delta=0.15$.

The results show that the time evolution is dominated by splitting of the initial wave into (at least) three main solitons. It seems that in long time evolution, one can expect more distinct emergence of the fourth one. In Fig.~\ref{abd15_Bs3.}, one can notice the increase of the amplitude of the highest soliton during its motion over the bottom bump, which is almost unnoticeable in Fig.~\ref{abd15_Ks3.}.

\begin{figure}[bht] \begin{center} \resizebox{0.9\columnwidth}{!}{\includegraphics{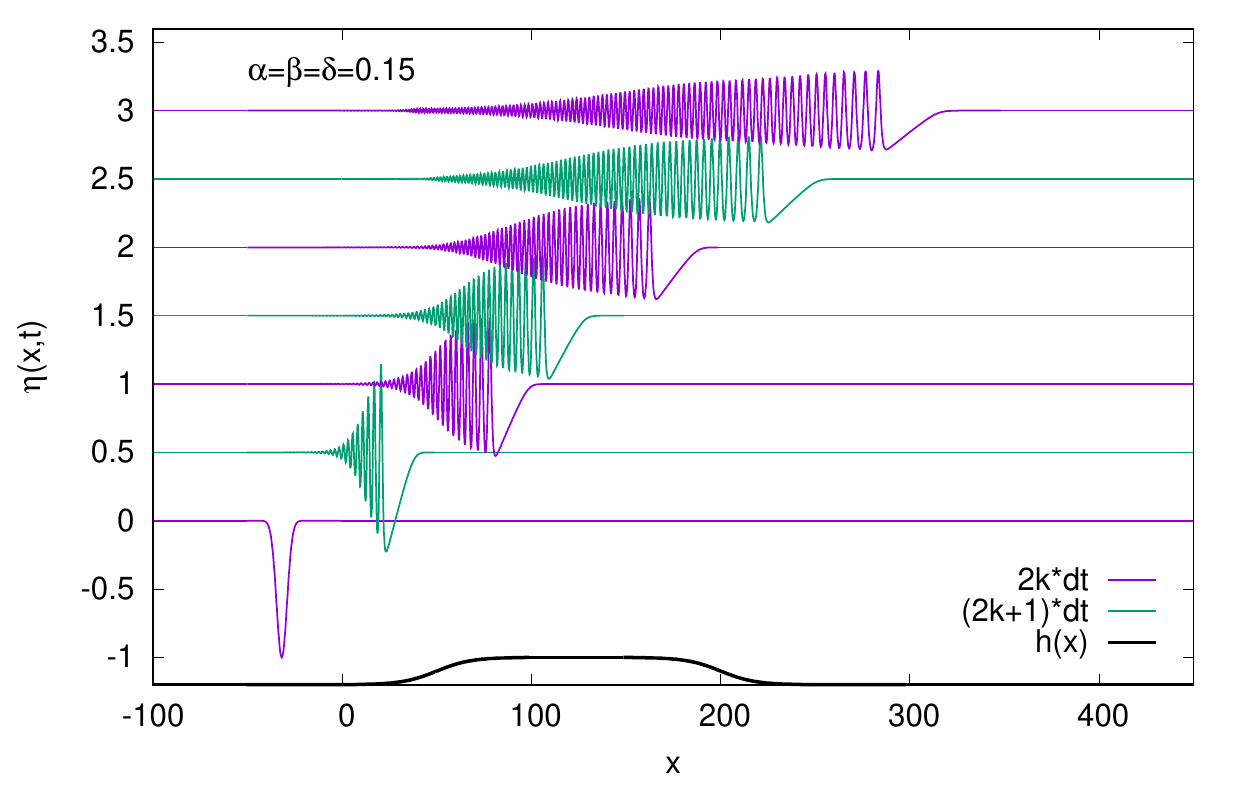}} \caption{Time evolution obtained according to the KdV equation (\ref{kdvD}). Initial Gaussian profile with the triple volume of the KdV soliton, the same velocity, but the inverse amplitude.  Here, time step between the consecutive profiles is $dt=64$.} 
\label{abd15_Ks3i.}\end{center} \end{figure}

\begin{figure}[bht] \begin{center} \resizebox{0.9\columnwidth}{!}{\includegraphics{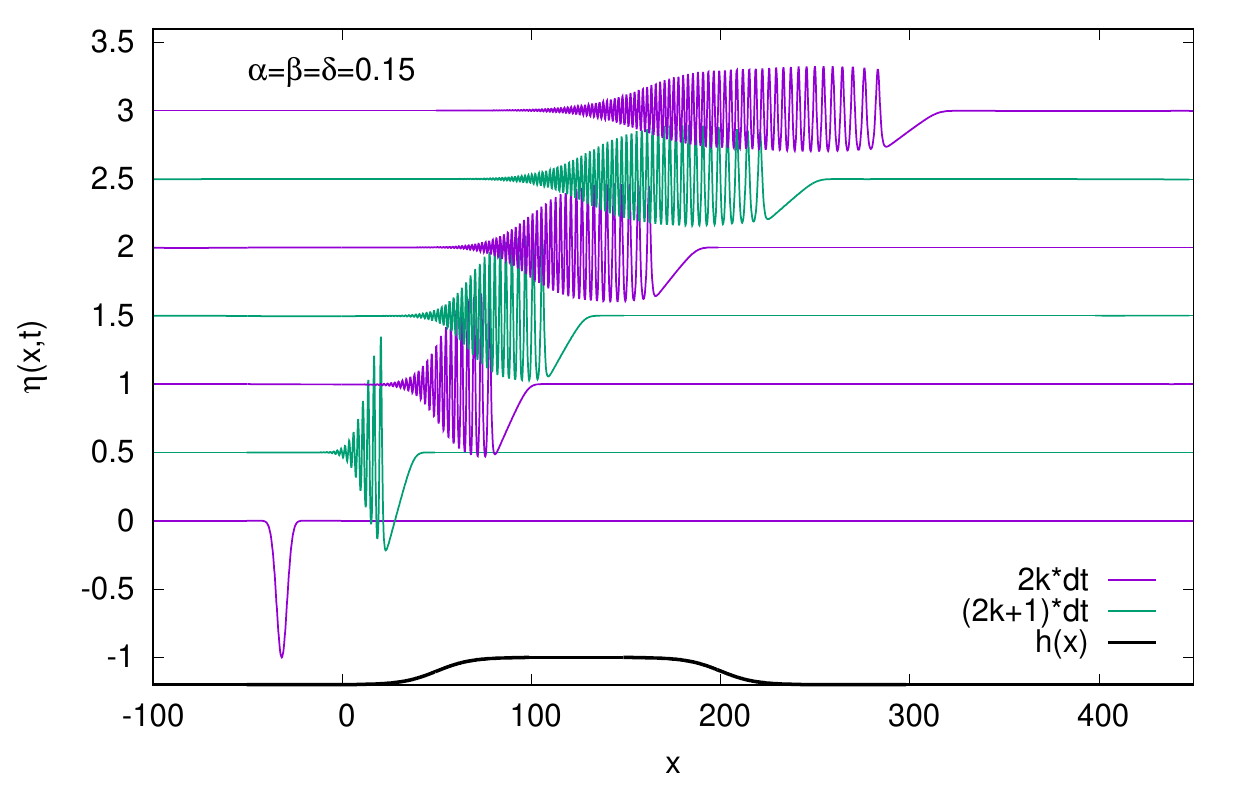}} \caption{Time evolution obtained according to Boussinesq's equations (\ref{4hx1})-(\ref{5hx1}). Initial Gaussian profile with the triple volume of the KdV soliton, the same velocity, but the inverse amplitude.  Here, time step between the consecutive profiles is $dt=64$. } 
\label{abd15_Bs3i.}\end{center} \end{figure}

In Figs.~\ref{abd15_Ks3i.} and \ref{abd15_Bs3i.} we present the cases of the time evolution with equation parameters as in Figs.~\ref{abd15_Ks3.} and \ref{abd15_Bs3.} but assuming that the initial distortion has an inverse form than the appropriate soliton (depression instead elevation). In these cases, the waves behave in an entirely different way.

\textcolor{blue}{The cases with $\alpha=\beta=\delta=0.25$ with the inverse initial wave profile (not shown here) suggest {\bf chaotic dynamics}.}

\subsection{KdV2 case}\label{kdv2-inv}

\begin{figure}[bht] \begin{center} \resizebox{0.9\columnwidth}{!}{\includegraphics{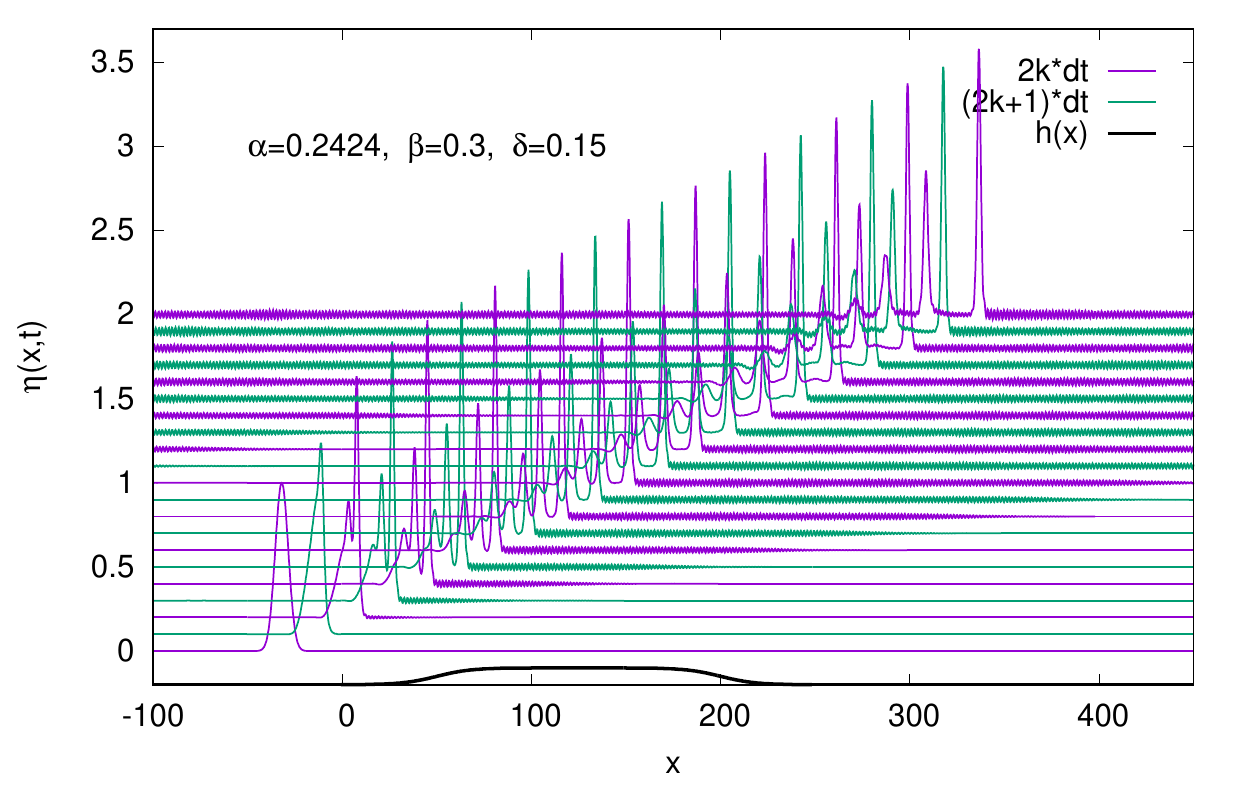}} \caption{Time evolution obtained according to the KdV2 equation (\ref{nkdv2d}). Initial Gaussian profile with the triple volume of the KdV2 soliton, the same velocity and amplitude.  Here, time step between the consecutive profiles is $dt=16$.} \label{a24d15_K2s3.}\end{center} \end{figure}

\begin{figure}[bht] \begin{center} \resizebox{0.9\columnwidth}{!}{\includegraphics{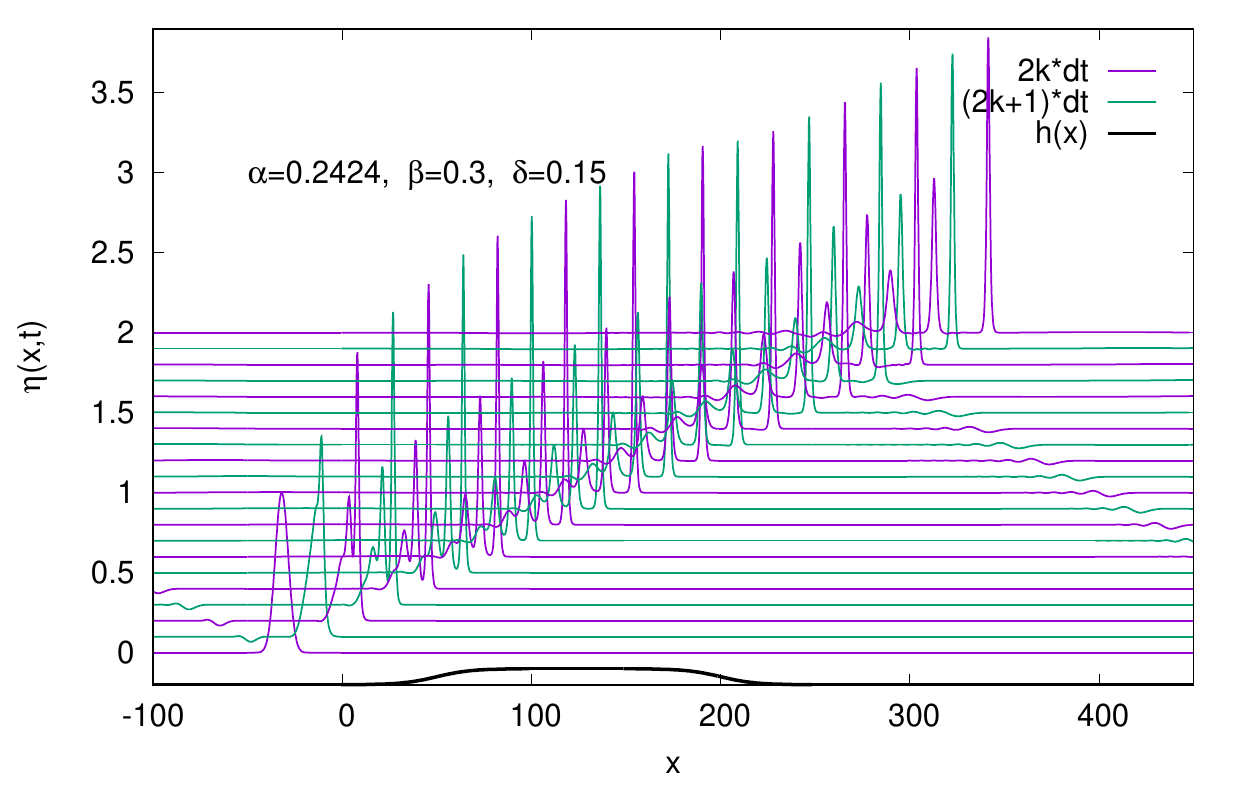}} \caption{Time evolution obtained according to Boussinesq's equations (\ref{4hx})-(\ref{5hx}). Initial Gaussian profile with the triple volume of the KdV2 soliton, the same velocity and amplitude.  Here, time step between the consecutive profiles is $dt=16$. } \label{a24d15_B2s3.}\end{center} \end{figure}

In Figs.~\ref{a24d15_K2s3.} and \ref{a24d15_B2s3.} we show the profiles of the time evolution of waves calculated according to equations (\ref{nkdv2d}) (KdV2 generalized for an uneven bottom) and (\ref{4hx})-(\ref{5hx})  (the corresponding Boussinesq equations), respectively. In both cases, the initial condition was taken as the Gaussian profile moving with the KdV soliton's velocity, the same amplitude, but with the triple volume of the fluid distortion from equilibrium. The parameters of wave equations are $\alpha=\beta=\delta=0.15$. Since the equations describe the macroscopic shallow water case, the parameter $\tau$ is set equal to zero.

The results displayed in Figs.~\ref{a24d15_K2s3.} and \ref{a24d15_B2s3.} show that the time evolution is dominated by splitting of the initial wave into (at least) four  solitons. It seems that in long time evolution, one can expect more distinct emergence of the fifth one. In Fig.~\ref{a24d15_K2s3.}, this splitting is accompanying by forwarding radiation of fast oscillations with tiny amplitude (the effect which also appeared in our earlier papers \cite{KRI,KRbook,RRIK}). 
In Fig.~\ref{a24d15_B2s3.}, one can notice the increase of the amplitude of the highest soliton during its motion over the bottom bump, which is difficult to see in Fig.~\ref{a24d15_K2s3.}.

In next Figs.~\ref{Ks3i_a24b3d15.} and \ref{Bs3i_a24b3d15.} we present the time evolution with the same parameters as those in Figs.~\ref{a24d15_K2s3.} and \ref{a24d15_B2s3.}. The only difference is that now the initial condition is taken as inverse of that in Figs.~\ref{a24d15_K2s3.} and \ref{a24d15_B2s3.}. This means that the initial condition has the form of depression instead elevation (normal for KdV2 equation).
Surprisingly, time evolution obtained directly from the generalized KdV2 equation (\ref{nkdv2d}) displayed in Fig.~\ref{Ks3i_a24b3d15.} differs substantially from the time evolution obtained from the appropriate Boussinesq's equations (\ref{4hx})-(\ref{5hx}). The time evolution of $w$ function, presented additionally in Fig.~\ref{BWs3i_a24b3d15.} is qualitatively very similar to the evolution of $\eta$ function. In contrast to these results obtained from the Boussinesq's equations, the time evolution resulting from the generalized KdV2 equation  (\ref{nkdv2d}) look {\bf chaotic}. 
This behavior may have the following cause. The KdV2 equation is only one of those considered in this paper, whose analytical solution is the so-called  \emph{embedded soliton}. This point deserves further study.

\begin{figure}[bht] \begin{center} \resizebox{0.9\columnwidth}{!}{\includegraphics{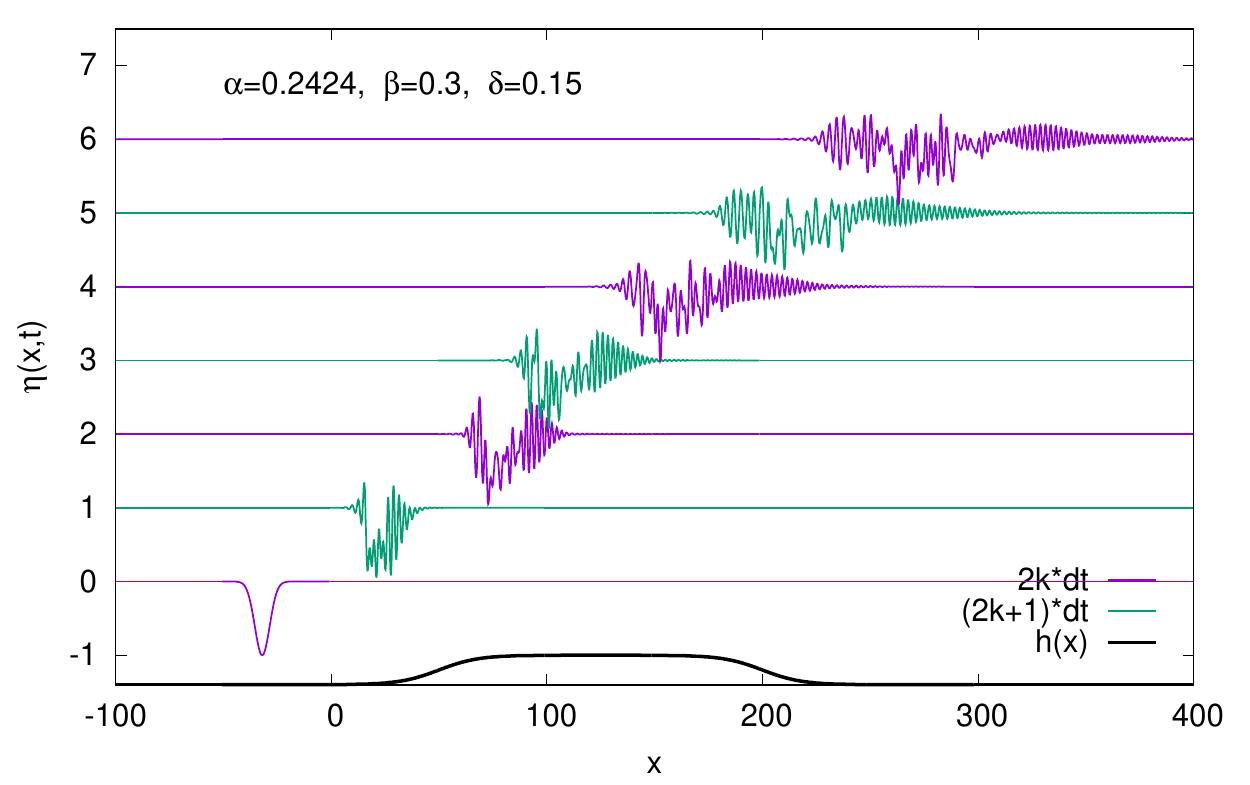}} \caption{Time evolution obtained according to the KdV2 equation (\ref{nkdv2d}). Initial Gaussian profile with the triple volume of the KdV2 soliton, the same velocity, but the inverse amplitude.  Here, time step between the consecutive profiles is $dt=64$.} \label{Ks3i_a24b3d15.}\end{center} \end{figure}

\begin{figure}[bht] \begin{center}
\resizebox{0.9\columnwidth}{!}{\includegraphics{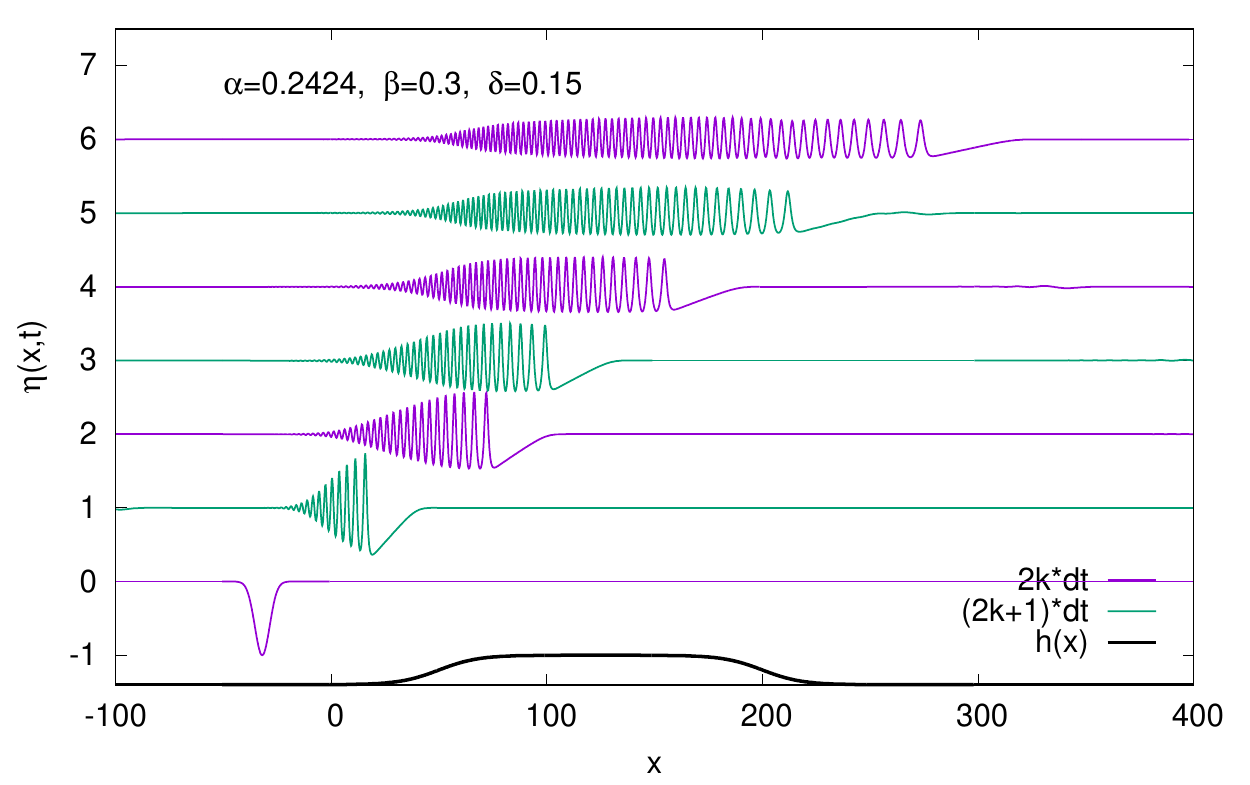}} \caption{Time evolution of the surface wave $\eta(x,t)$ obtained according to Boussinesq's equations (\ref{4hx})-(\ref{5hx}). 
 Initial Gaussian profile with the triple volume of the KdV2 soliton, the same velocity, but the inverse amplitude.  Here, time step between the consecutive profiles is $dt=64$. } \label{Bs3i_a24b3d15.}\end{center} \end{figure}

\begin{figure}[bht] \begin{center} \resizebox{0.9\columnwidth}{!}{\includegraphics{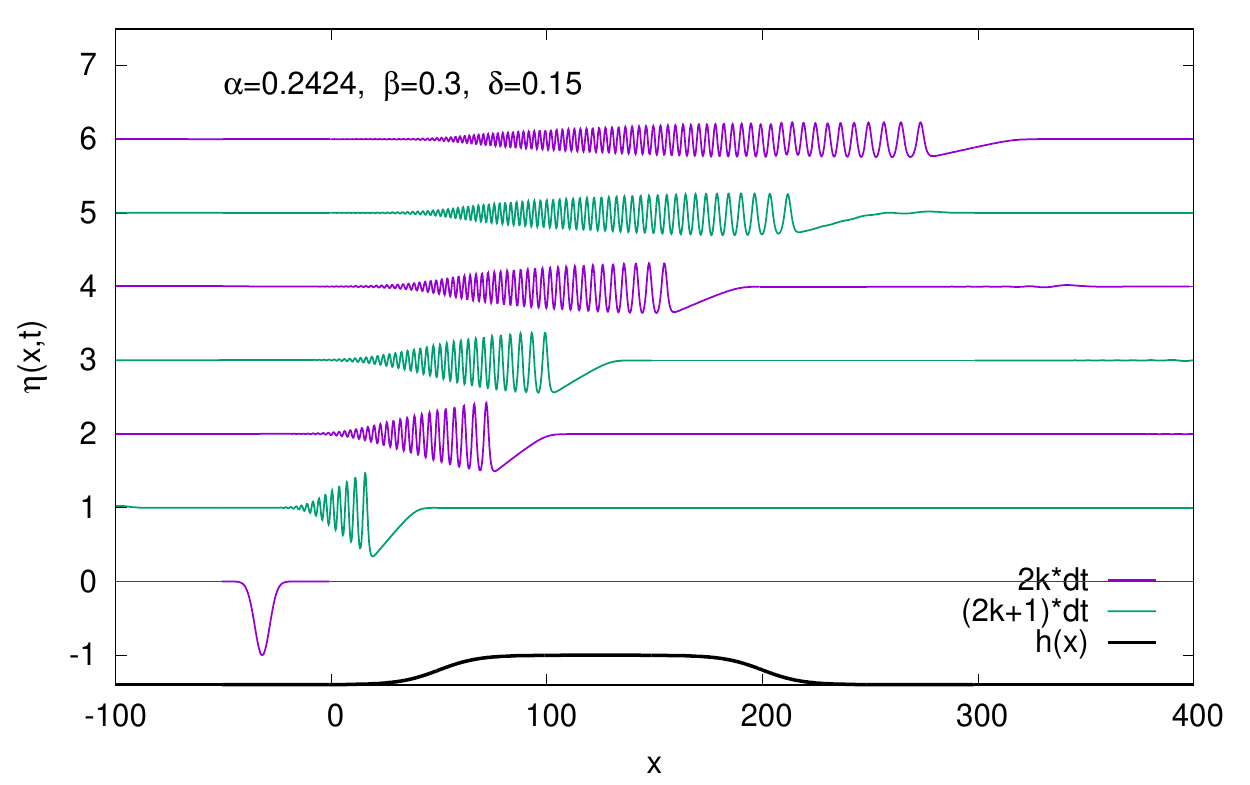}} \caption{Time evolution of the function $w(x,t)$ obtained according to Boussinesq's equations (\ref{4hx})-(\ref{5hx}). Initial Gaussian profile with the triple volume of the KdV2 soliton, the same velocity, but the inverse amplitude.  Here, time step between the consecutive profiles is $dt=64$. } 
\label{BWs3i_a24b3d15.}\end{center} \end{figure}

\subsection{5th-order KdV case}\label{kdv5-inv}

Let us recall, that analytic soliton solutions to the 5th-order KdV equation in the form $\eta(x,t)= A\,Sech[B(x-vt)]^{4}$ exist only when $\frac{1}{3} <\tau < \frac{2\sqrt{30}-5}{15}  \approx 0.39696$. Properties of wave motion when $\tau$ is close to $\frac{1}{3}$ and when  $\tau$ is close to $0.397$ differ substantially from each other. Therefore, we present examples of time evolution of waves describe by 5th-order KdV equation for two cases of $\tau$.

\subsubsection{Small $\tau$, close to lower limit} \label{5kdvt35}

\begin{figure}[bht] \begin{center}
\resizebox{0.9\columnwidth}{!}{\includegraphics[angle=270]{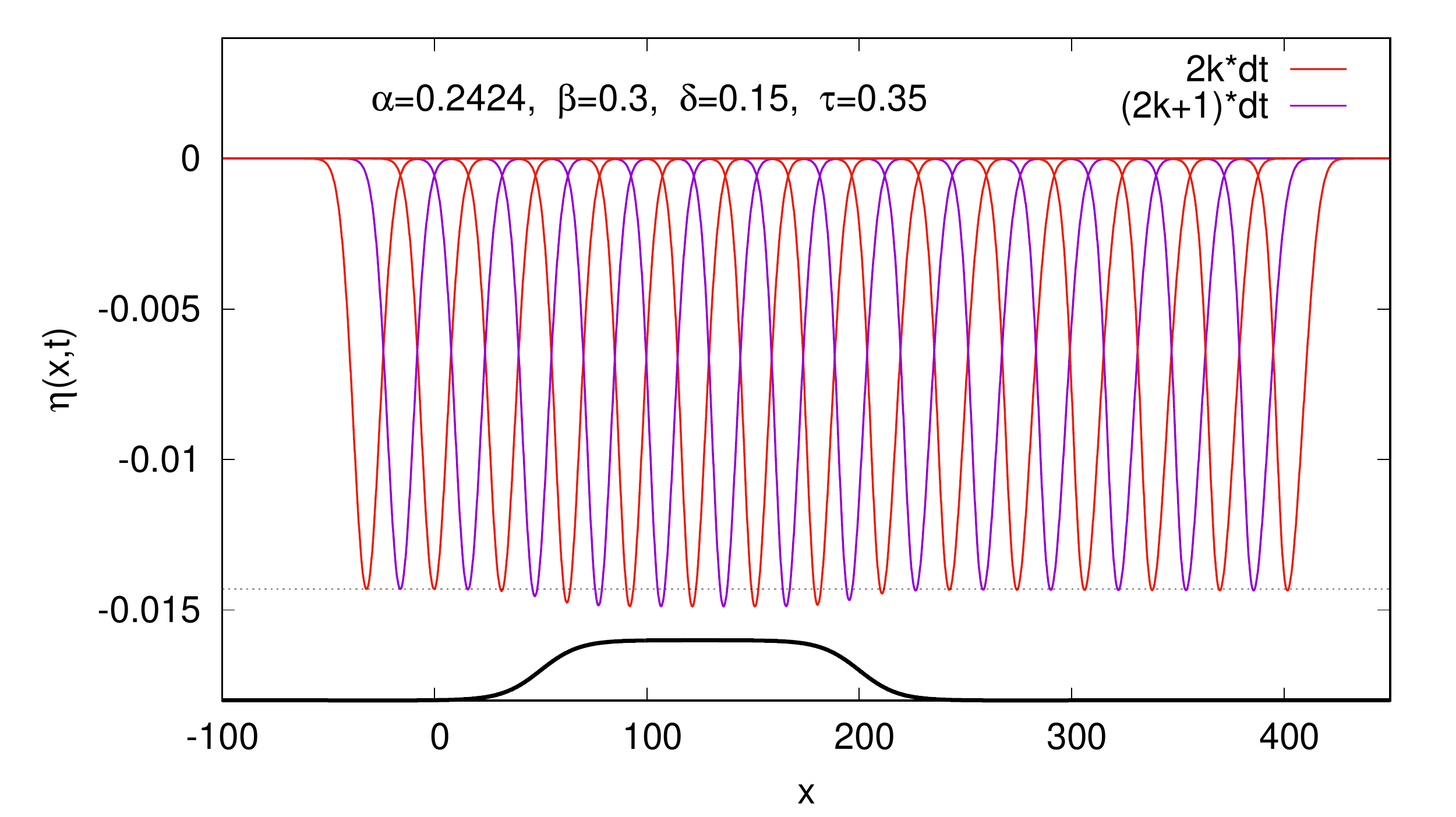}} \caption{Time evolution obtained according to the generalized 5th-order KdV equation (\ref{5kdvQ}). Initial Gaussian profile with the triple volume of the KdV5 soliton, the same velocity and amplitude.  Here, time step between the consecutive profiles is $dt=16$.} \label{5Ka24_Ks3.}\end{center} \end{figure}

\begin{figure}[bht] \begin{center} \resizebox{0.9\columnwidth}{!}{\includegraphics[angle=270]{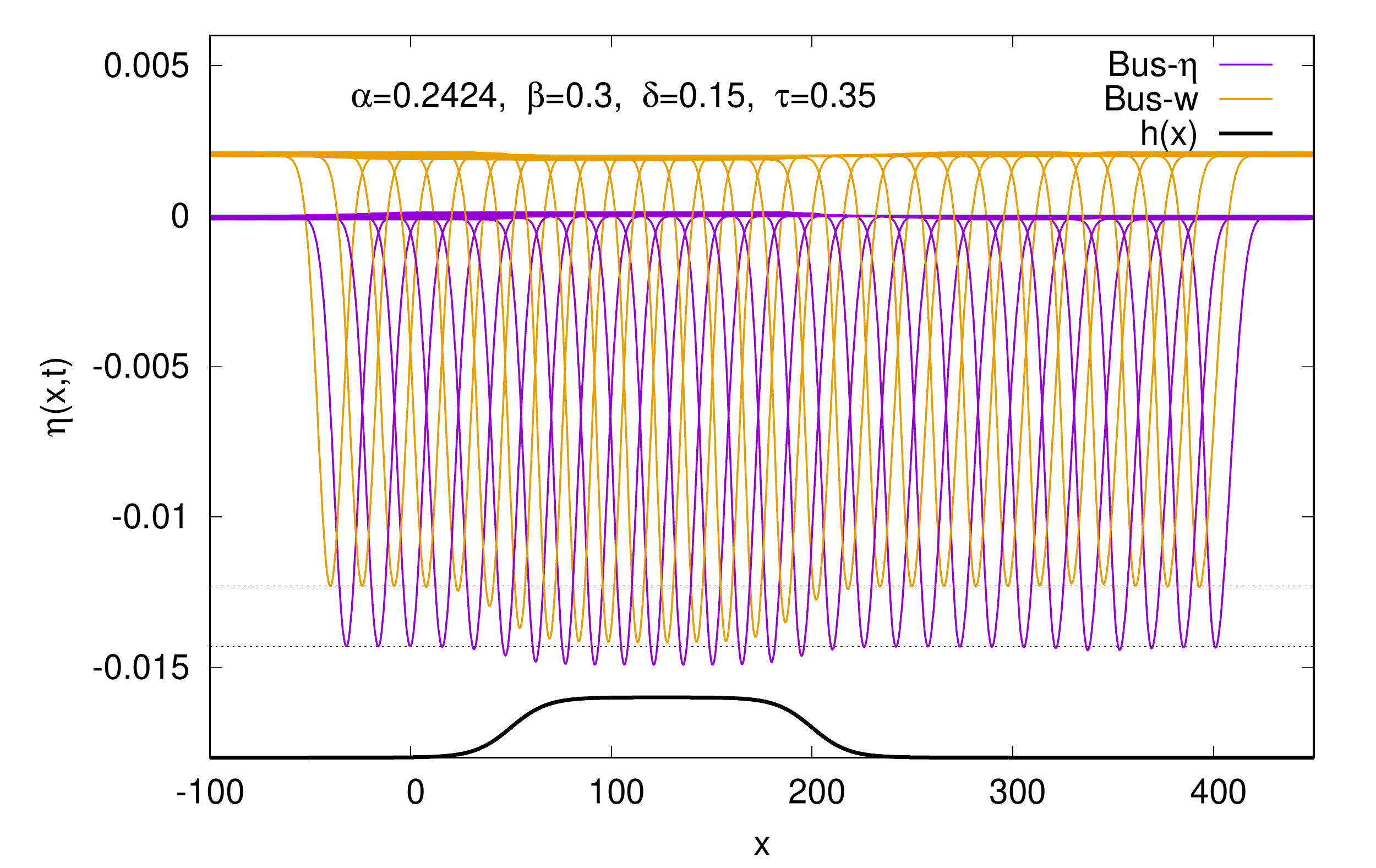}} \caption{Time evolution obtained according to Boussinesq's equations (\ref{3Ba2d2})-(\ref{4Ba2d2}). Initial Gaussian profile with the triple volume of the KdV5 soliton, the same velocity and amplitude.  Here, time step between the consecutive profiles is $dt=16$. } 
\label{5Ka24_BWs3.}\end{center} \end{figure}

Begin with $\tau=0.35$, as in \cite[Sec.~8]{KRcnsns}.
In Figs.~\ref{5Ka24_Ks3.} and \ref{5Ka24_BWs3.} we show the profiles of the time evolution of waves calculated according to equations (\ref{5kdvQ}) (5th-order KdV generalized for an uneven bottom) and (\ref{3Ba2d2})-(\ref{4Ba2d2}) (the corresponding Boussinesq equations), respectively. In both cases, the initial condition was taken as the Gaussian profile moving with the KdV5 soliton's velocity, the same amplitude, but with the triple volume of the fluid distortion from equilibrium. The parameters of wave equations are $\alpha=0.2424, \beta=0.3, \delta=0.15$.

Surprisingly, in this case, the wave profiles remain almost unchanged during the evolution, with only a slight increase of the amplitude when the wave travels over the bottom bump. As in most other cases, the impact of the varying bottom in the surface wave is more significant in Boussinesq's equations.

\begin{figure}[bht] \begin{center} \resizebox{0.9\columnwidth}{!}{\includegraphics[angle=270]{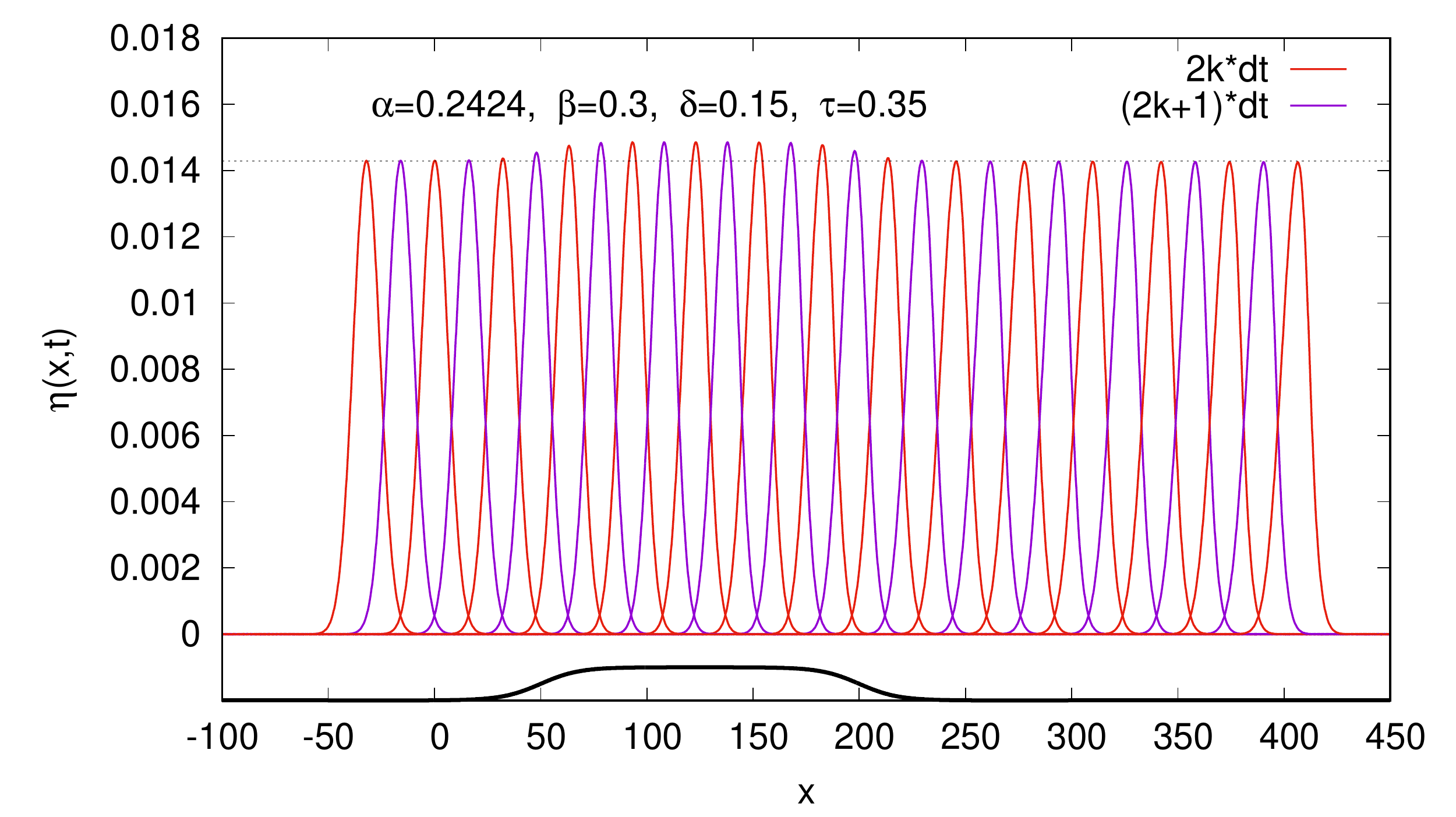}} \caption{Time evolution obtained according to the generalized 5th-order KdV equation (\ref{5kdvQ}). Initial Gaussian profile with the triple volume of the KdV5 soliton, the same velocity, but the inverse initial distortion.  Here, time step between the consecutive profiles is $dt=16$.} 
\label{5Ka24_Ks3i.}\end{center} \end{figure}

\begin{figure}[bht] \begin{center} \resizebox{0.9\columnwidth}{!}{\includegraphics[angle=270]{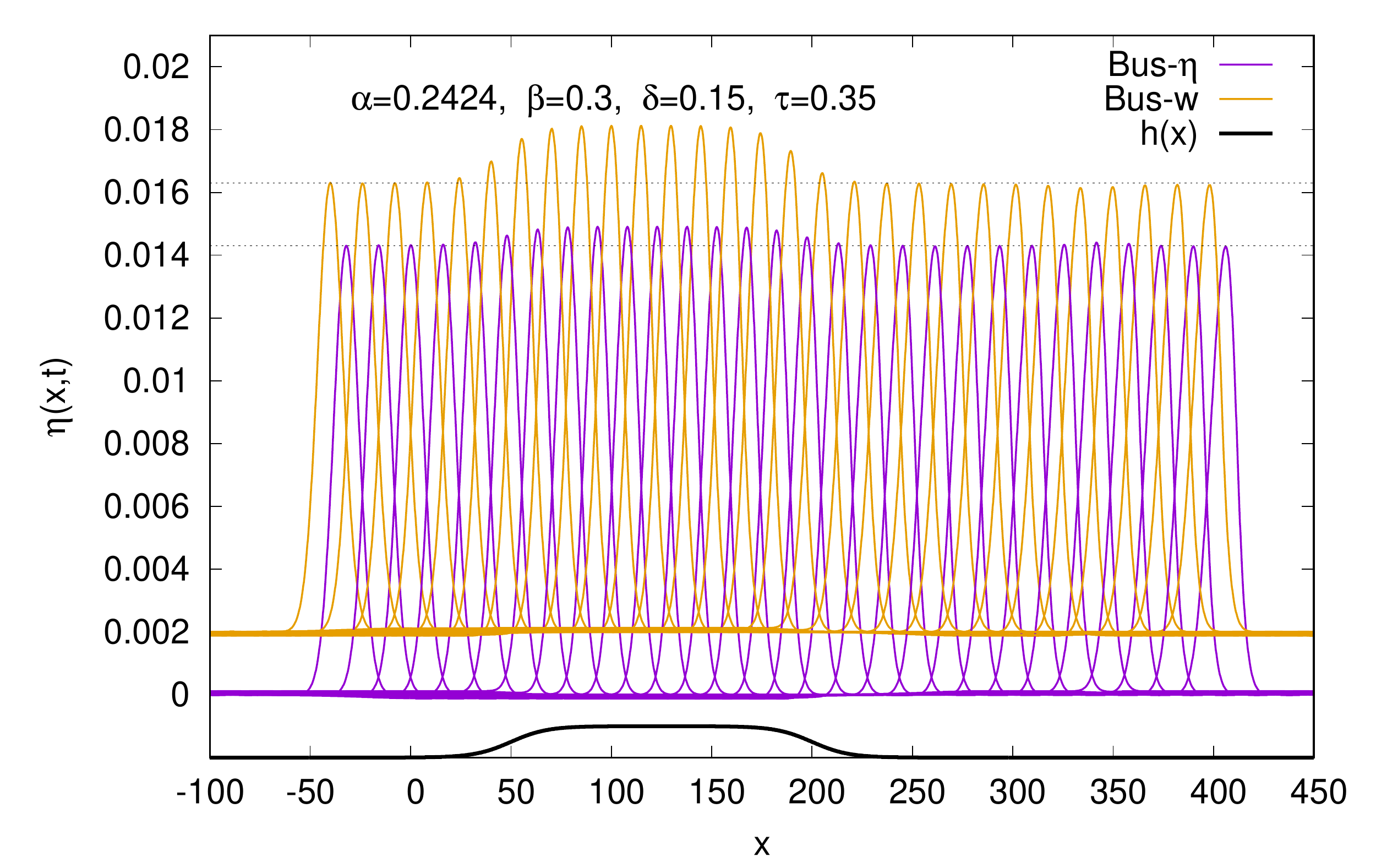}} \caption{Time evolution obtained according to Boussinesq's equations (\ref{3Ba2d2})-(\ref{4Ba2d2}). Initial Gaussian profile with the triple volume of the KdV5 soliton, the same velocity, but the inverse initial distortion. Here, time step between the consecutive profiles is $dt=16$. } 
\label{5Ka24_BWs3i.}\end{center} \end{figure}

In Figs.~\ref{5Ka24_Ks3i.} and \ref{5Ka24_BWs3i.}, we present the cases of the time evolution with equation parameters as in Figs.~\ref{5Ka24_Ks3.} and \ref{5Ka24_BWs3.}    but assuming that the initial distortion has an inverse form than the appropriate soliton (elevation instead of depression). It is again surprising that in these cases, the profiles look like inverted profiles shown in Figs.~\ref{5Ka24_Ks3.} and \ref{5Ka24_BWs3.}.


\subsubsection{Large $\tau$, close to upper limit} \label{t38}

\begin{figure}[bht] \begin{center} \resizebox{0.9\columnwidth}{!}{\includegraphics[angle=270]{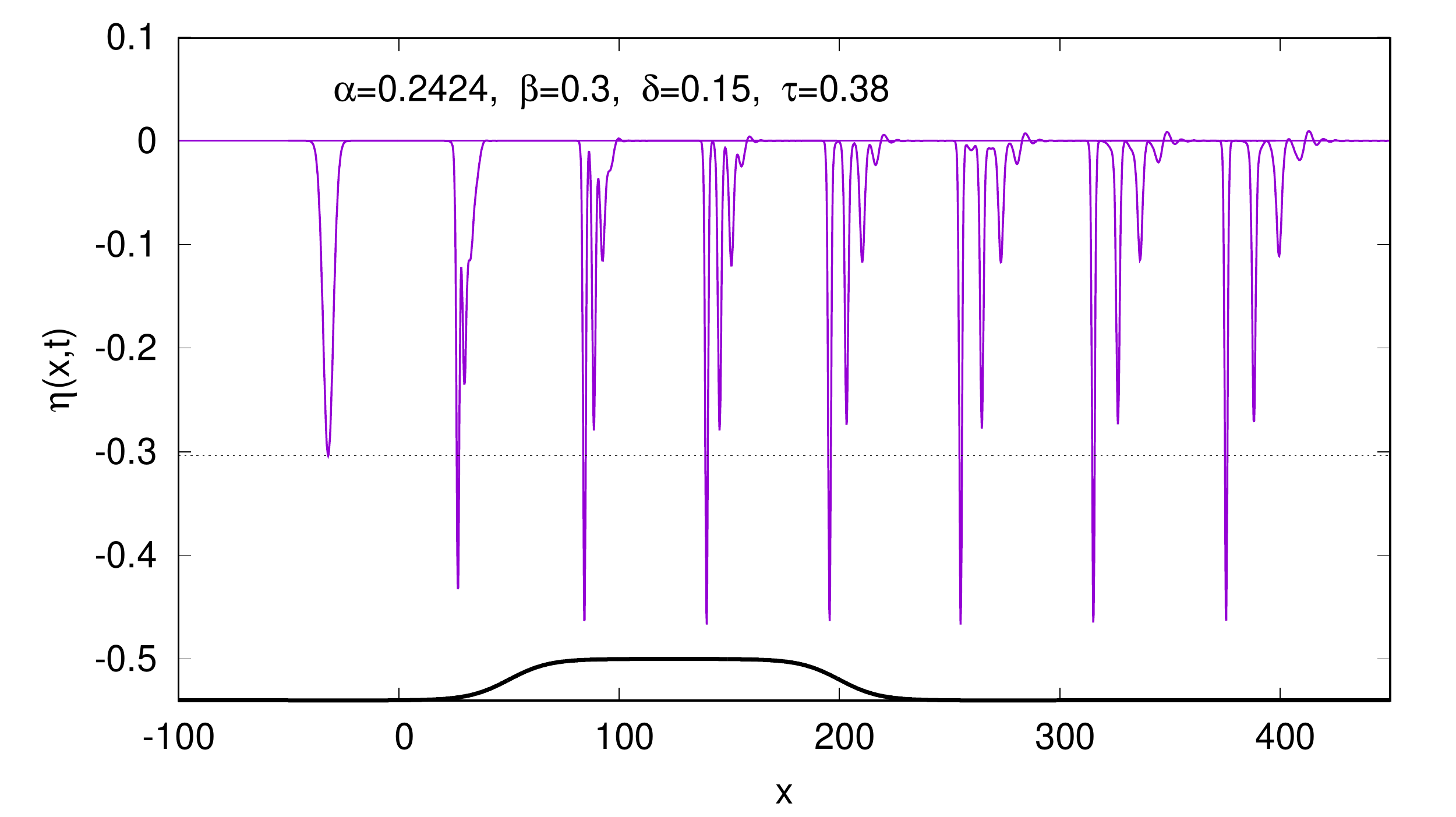}}
 \caption{Time evolution obtained according to the generalized 5th-order KdV equation (\ref{5kdvQ}). Initial Gaussian profile with the triple volume of the KdV5 soliton, the same velocity and amplitude.  Here, time step between the consecutive profiles is $dt=64$.} 
\label{5Ka24t38_Ks3.}\end{center} \end{figure}

\begin{figure}[bht] \begin{center} \resizebox{0.9\columnwidth}{!}{\includegraphics[angle=270]{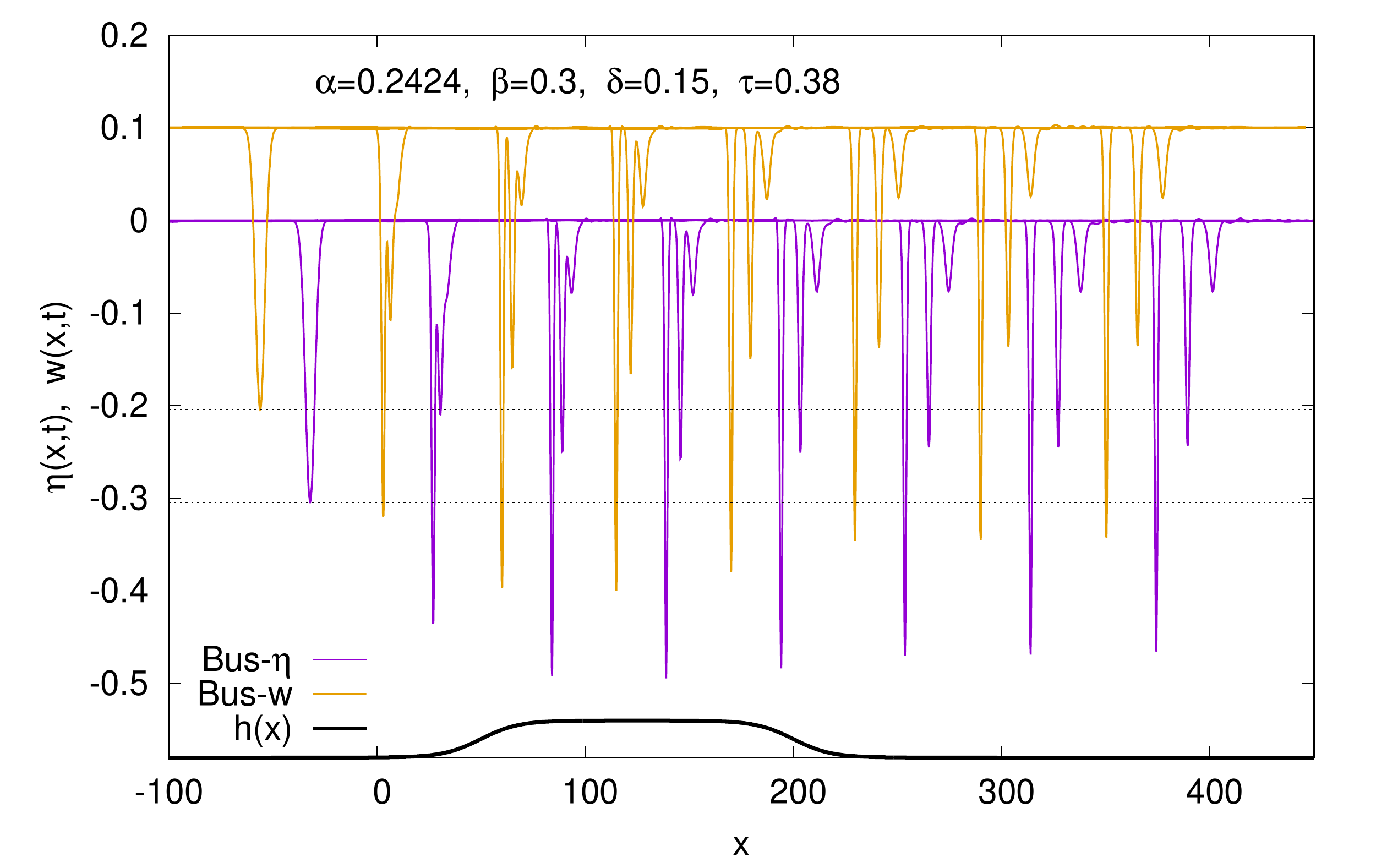}}
 \caption{Time evolution obtained according to Boussinesq's equations (\ref{3Ba2d2})-(\ref{4Ba2d2}). Initial Gaussian profile with the triple volume of the KdV5 soliton, the same velocity and amplitude.  Here, time step between the consecutive profiles is $dt=64$. } 
\label{5Ka24t38_BWs3.}\end{center} \end{figure}

Now, we use $\tau=0.38$. Figures \ref{5Ka24t38_Ks3.} and \ref{5Ka24t38_BWs3.} present analogous time evolution as Figs.~\ref{5Ka24_Ks3.} and \ref{5Ka24_BWs3.}. To avoid profile overlaps we displayed profiles at larger time intervals $dt=64$.
The profiles of $w$ function are shifted by 32 left and by 0.1 up. 
It is clear that in these cases the time evolution is dominated by the process of splitting of the initial wave into at least four solitons (during the time of calculation). Results obtained with single equation (\ref{5kdvQ}) and Boussinesq's equations (\ref{3Ba2d2})-(\ref{4Ba2d2}) are very similar. In the former case the impact of the bottom bump is almost unnoticeable, in the latter is visible but also small.

\begin{figure}[bht] \begin{center} \resizebox{0.9\columnwidth}{!}{\includegraphics[angle=270]{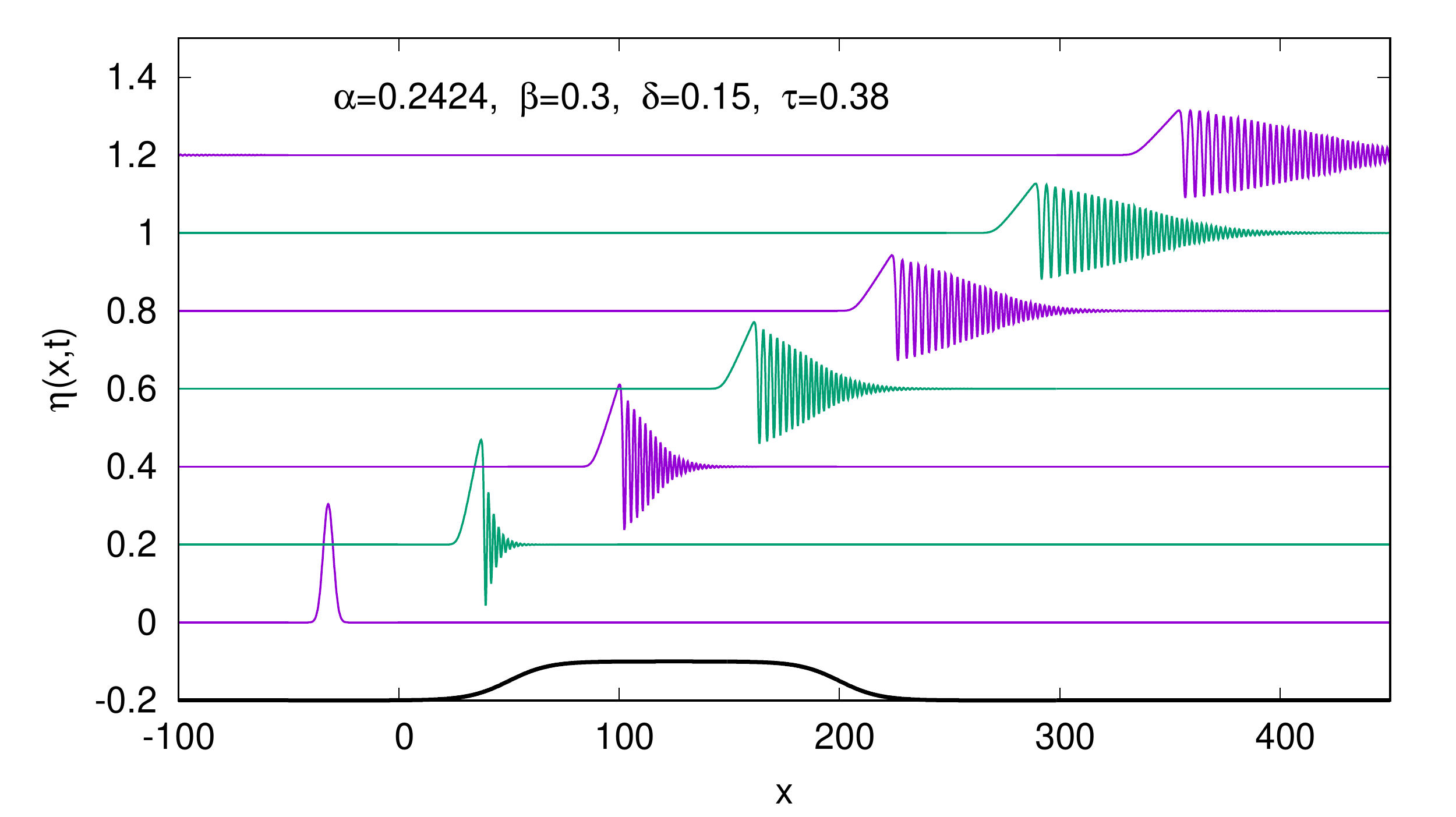}}
 \caption{Time evolution obtained according to the generalized 5th-order KdV equation  (\ref{5kdvQ}). Initial Gaussian profile with the triple volume of the KdV5 soliton, the same velocity, but the inverse initial distortion.  Here, time step between the consecutive profiles is $dt=64$.} 
\label{5Ka24t38_Ks3i.}\end{center} \end{figure}

\begin{figure}[bht] \begin{center} \resizebox{0.9\columnwidth}{!}{\includegraphics[angle=270]{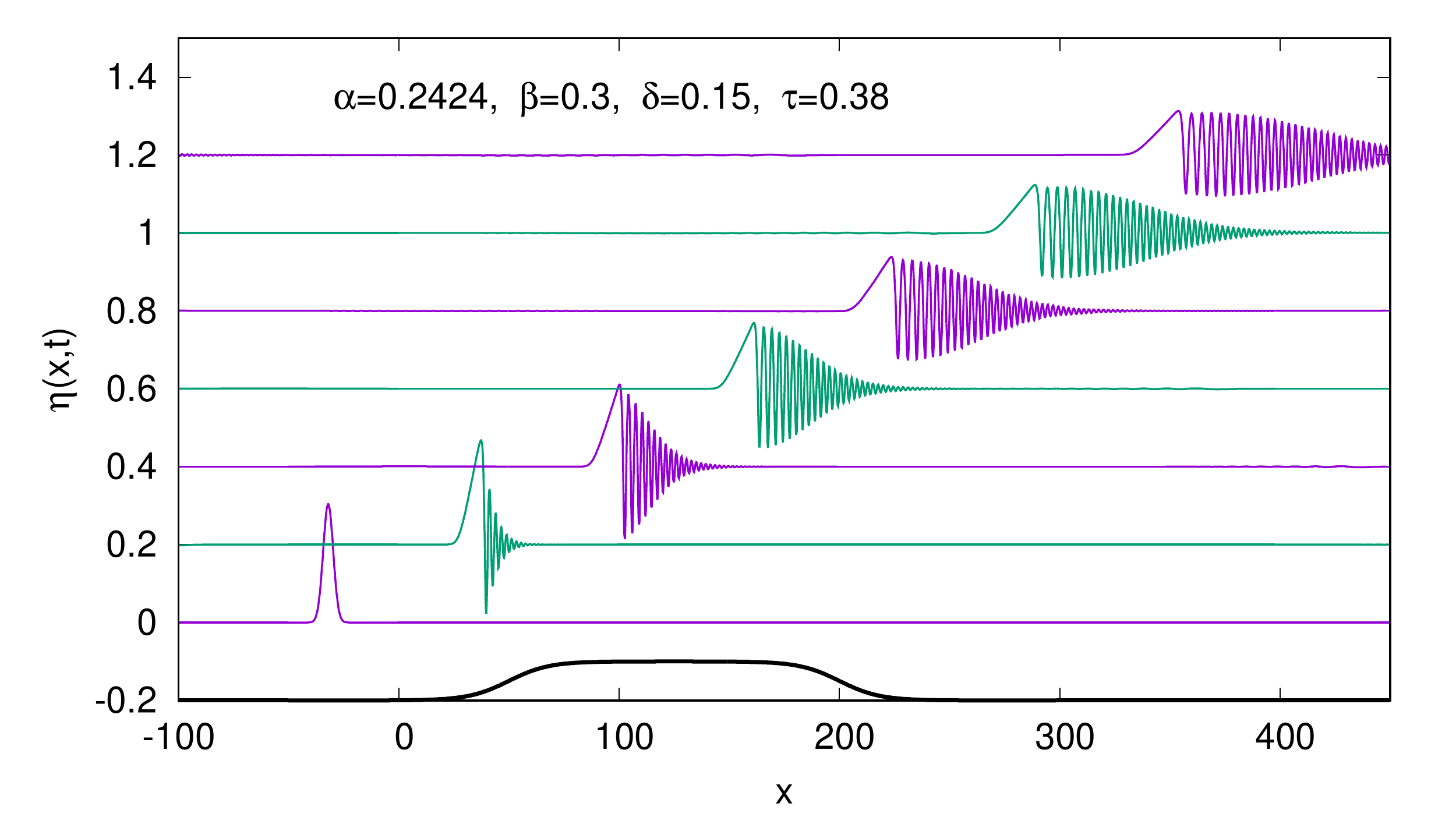}}
 \caption{Time evolution obtained according to Boussinesq's equations (\ref{3Ba2d2})-(\ref{4Ba2d2}). Initial Gaussian profile with the triple volume of the KdV5 soliton, the same velocity, but the inverse initial distortion. Here, time step between the consecutive profiles is $dt=64$. } 
\label{5Ka24t38_BWs3i.}\end{center} \end{figure}

In Figs.~\ref{5Ka24t38_Ks3i.} and \ref{5Ka24t38_BWs3i.} we present cases analogous to those shown in Figs.~\ref{5Ka24_Ks3i.} and \ref{5Ka24_BWs3i.} but for $\tau=0.38$. The initial condition is taken as the Gaussian with the volume three times greater than the volume of the soliton. However, the initial condition is inverse than the 'normal' one. It is the elevation instead of the depression. 
In Fig.~\ref{5Ka24t38_BWs3i.} only $\eta$ function is displayed.
In these cases, the behavior of the wave evolution is qualitatively similar to corresponding cases (with inverse initial conditions) for different equations.


\subsection{Gardner equation}\label{GE-inv}

\subsubsection{Case corresponding to shallow water $(\tau=0)$}

In Figs.~\ref{a3bd09_Gs3.} and \ref{a3bd09_GBs3.}, we show the profiles of the time evolution of waves calculated according to equations (\ref{GardH}) (the Gardner equation generalized for an uneven bottom) and  (\ref{bbu3})-(\ref{bbu4}) (the corresponding Boussinesq equations), respectively. In both cases, the initial condition was taken as the Gaussian profile moving with the KdV soliton's velocity, the same amplitude, but with the triple volume of the fluid distortion from equilibrium. The parameters of wave equations are $\alpha=0.3, \beta=\delta=0.09$. Since the equations  describe the macroscopic shallow water case, the parameter $\tau$ is set equal to zero. The parameter $\Delta=1$ is chosen for the Gardner soliton. 

The results displayed in Figs.~\ref{a3bd09_Gs3.} and \ref{a3bd09_GBs3.} show that in these cases, the time evolution is dominated by splitting of the initial wave into (at least) two solitons. The last displayed profiles suggest that in long time evolution one can expect more distinct emergence of the third one. This property is slightly better pronounced in Fig.~\ref{a3bd09_GBs3.}, but the time evolution of the surface wave is almost the same for both figures.

In next Figs.~\ref{a3bd09t0_Gi3.} and \ref{a3bd09t0_GBi3.}, we present the time evolution with the same parameters as those in Figs.~\ref{a3bd09_Gs3.} and \ref{a3bd09_GBs3.}. The only difference is that now the initial condition is taken as inverse of that in Figs.~\ref{a3bd09_Gs3.} and \ref{a3bd09_GBs3.}. This means that the initial condition has the form of depression instead of elevation (normal for shallow water case). The time evolution shown in these figures is entirely different from when initial displacement has a 'normal' sign. On the other hand, results obtained from the generalized Gardner equation and the corresponding  Boussinesq's system are almost identical. 

\subsubsection{Case corresponding to thin fluid layers $(\tau >\frac{1}{3})$}

When $(\tau >\frac{1}{3})$ surface tension plays an important role. Such a situation appears when the thickness of the fluid layer is very small. In the following examples, we set $\tau=1$. 

In Figs.~\ref{a3bd09t1_Gs3a.} and \ref{a3bd09t1_GBs3a.}, we show the profiles of the time evolution of waves calculated according to equations  (\ref{GardH}) (the Gardner equation generalized for an uneven bottom) and   (\ref{bbu3})-(\ref{bbu4}) (the corresponding Boussinesq equations), respectively. In both cases, the initial condition was taken as the Gaussian profile moving with the Gardner soliton's velocity, the same amplitude, but with the triple volume of the fluid distortion from equilibrium. The parameters of wave equations are $\alpha=0.3, \beta=\delta=0.09$. The parameter $\Delta=1$ is chosen for the Gardner soliton. 

\begin{figure}[bht] \begin{center} \resizebox{0.9\columnwidth}{!}{\includegraphics{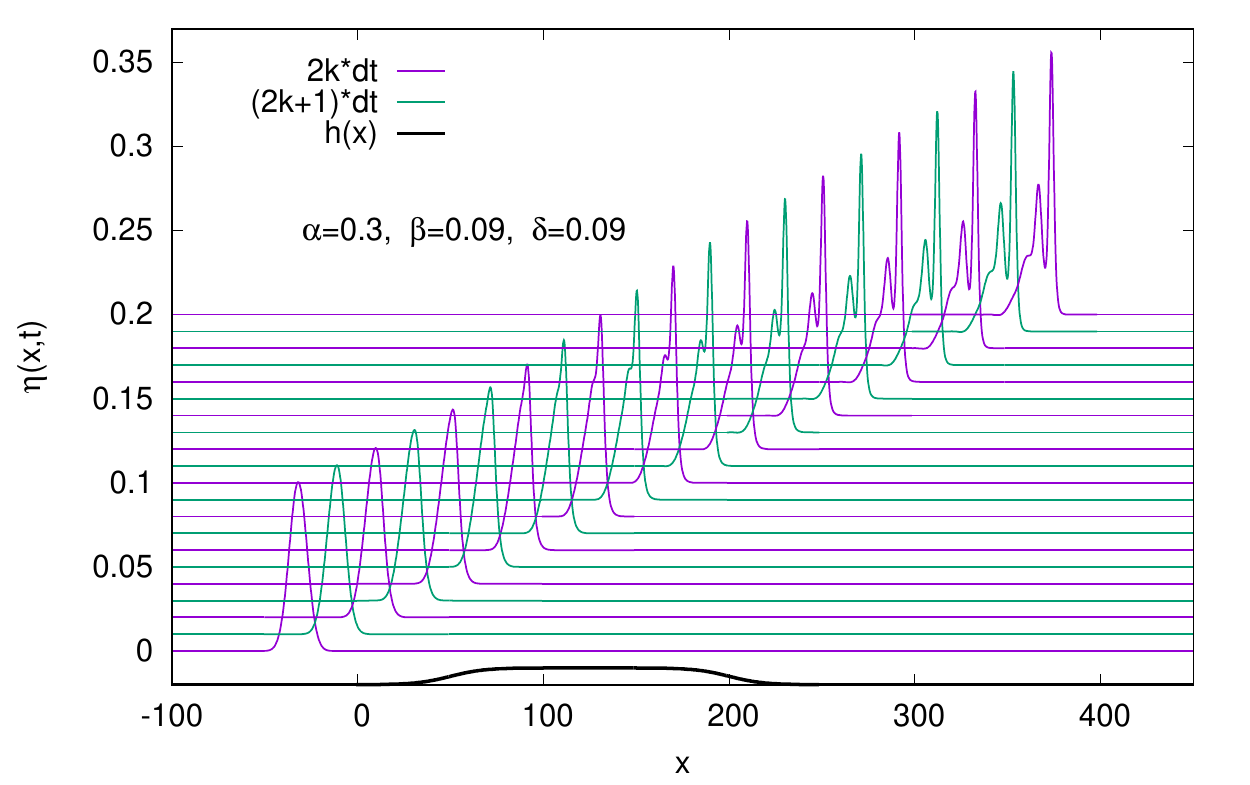}} \caption{Time evolution obtained according to the Gardner equation generalized for the uneven bottom (\ref{GardH}) with $\tau=0$. Initial Gaussian profile with the triple volume of the Gardner soliton, the same amplitude, and velocity. $dt=16$.} \label{a3bd09_Gs3.}\end{center} \end{figure}

\begin{figure}[bht] \begin{center}
\resizebox{0.9\columnwidth}{!}{\includegraphics{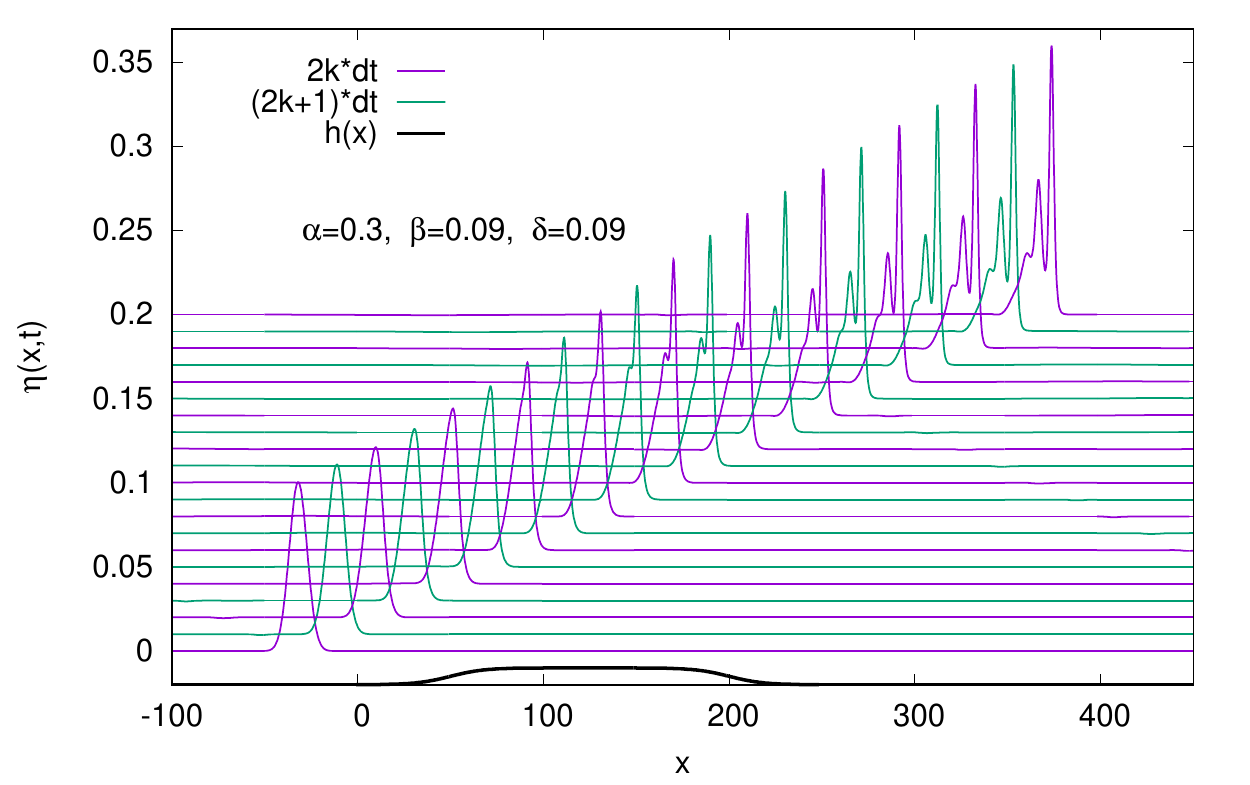}} \caption{Time evolution obtained according to Boussinesq's equations, generalized for the uneven bottom (\ref{bbu3})-(\ref{bbu4}) with $\tau=0$. Initial Gaussian profile with the the triple volume of the Gardner soliton, the same amplitude, and velocity.  Only surface wave $\eta(x,t)$ is displayed. $dt=16$.} 
\label{a3bd09_GBs3.}\end{center} \end{figure}

Similarly, as in Figs.~\ref{a3bd09_Gs3.} and \ref{a3bd09_GBs3.}, the time evolution is dominated by the splitting of the initial wave into several solitons, at least three. The fourth one seems to emerge  in the last calculated profiles, as well.
Here, the lowest solitons move faster than the higher ones, contrary to usual cases.
Similarly, as with $\tau=0$, the results obtained with the Gardner equation and the corresponding Boussinesq equations are almost the same. In the latter case, the impact of the bottom bump is slightly more pronounced. 

In Figs.~\ref{a3bd09t1_Gi3.} and \ref{a3bd09t1_GBi3.}, we used the same parameters as in Figs.~\ref{a3bd09t1_Gs3a.} and \ref{a3bd09t1_GBs3a.}, reversing only the sign of the initial displacement. The initial condition is then the elevation instead of depression. Again, the results obtained with the Gardner equation and the  corresponding Boussinesq equations are almost the same. They are, however, entirely different from those in Figs.~\ref{a3bd09t1_Gs3a.} and \ref{a3bd09t1_GBs3a.}.

\begin{figure}[bht] \begin{center} \resizebox{0.9\columnwidth}{!}{\includegraphics{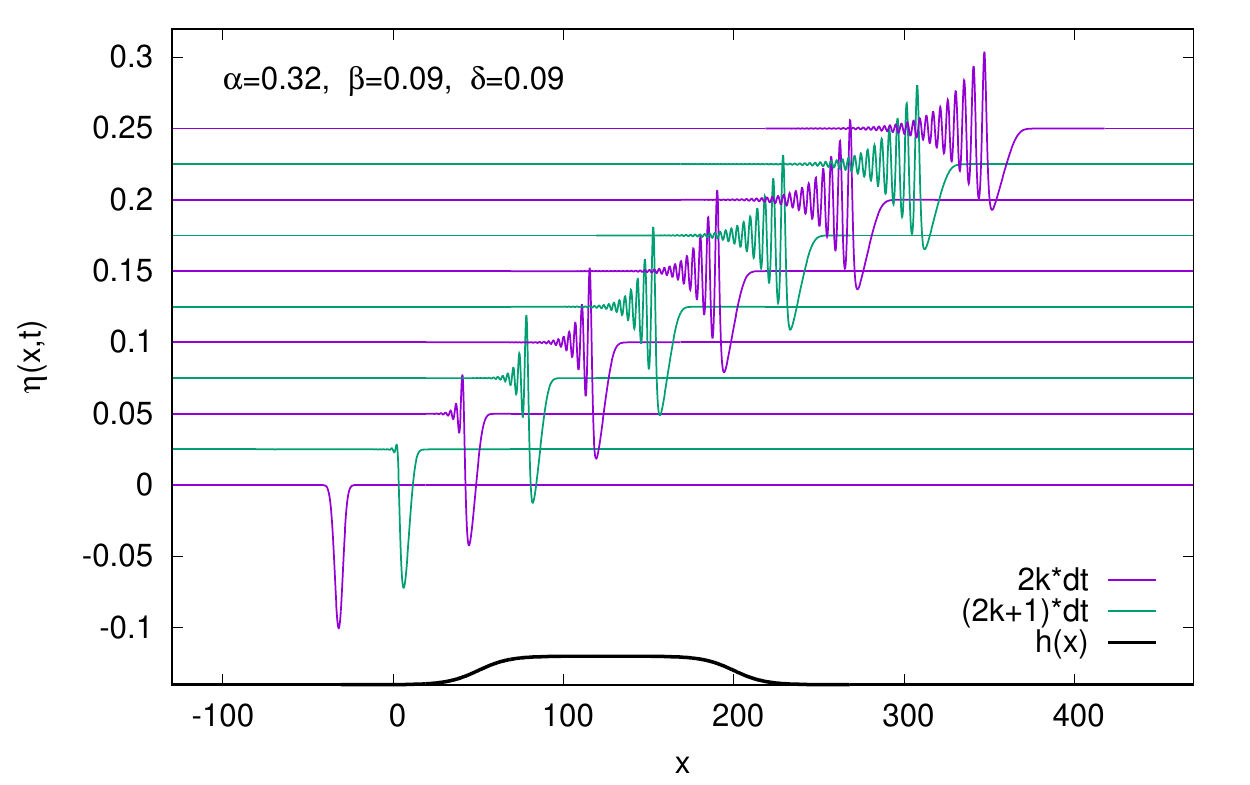}} \caption{Time evolution obtained according to the Gardner equation (\ref{GardH}) with $\tau=0$. Initial Gaussian profile, representing an elevation, with the triple volume of the Gardner soliton, the same velocity, but the inverse amplitude.  $dt=32$.} 
\label{a3bd09t0_Gi3.}\end{center} \end{figure}

\begin{figure}[bht] \begin{center} \resizebox{0.9\columnwidth}{!}{\includegraphics{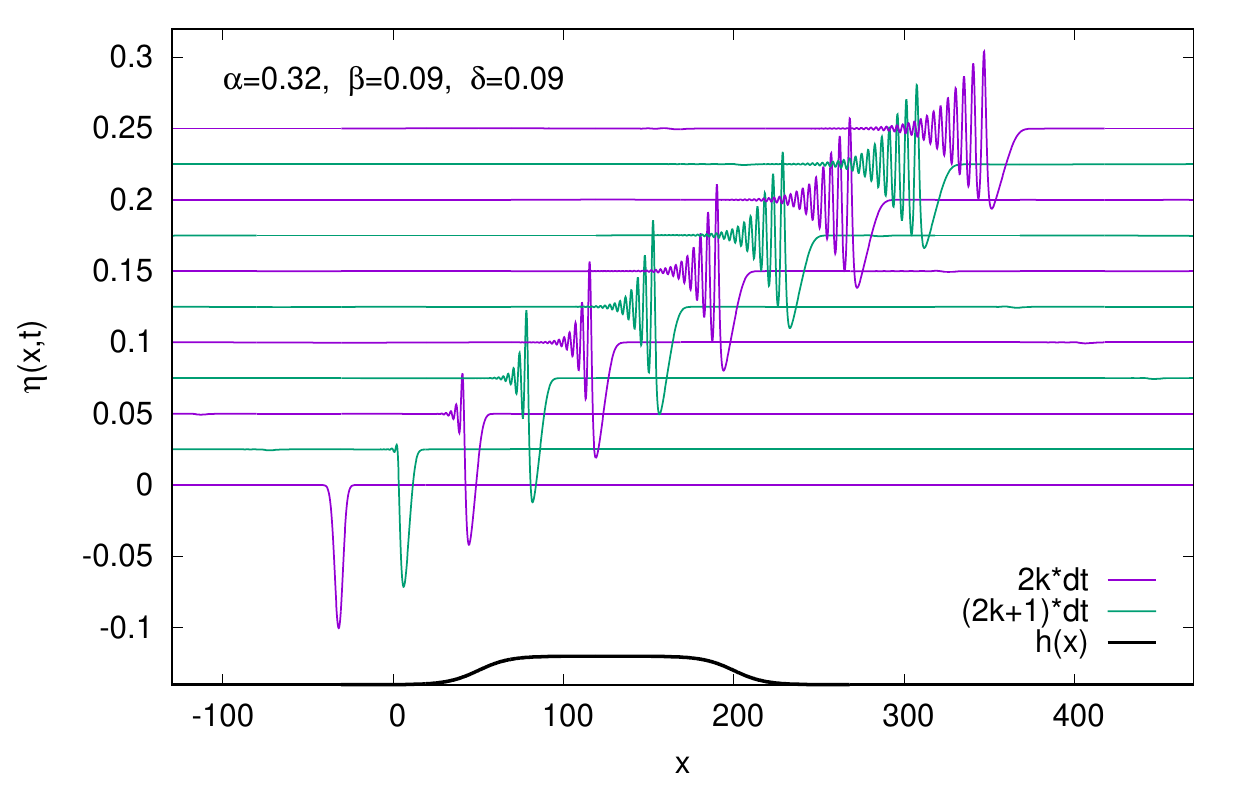}} \caption{Time evolution obtained according to Boussinesq's equations (\ref{bbu3})-(\ref{bbu4}) with $\tau=1$. Initial Gaussian profile, representing an elevation, with the the triple volume of Gardner soliton, the same velocity, but the inverse amplitude.  $dt=32$.} 
\label{a3bd09t0_GBi3.}\end{center} \end{figure}

\begin{figure}[bht] \begin{center} \resizebox{0.9\columnwidth}{!}{\includegraphics{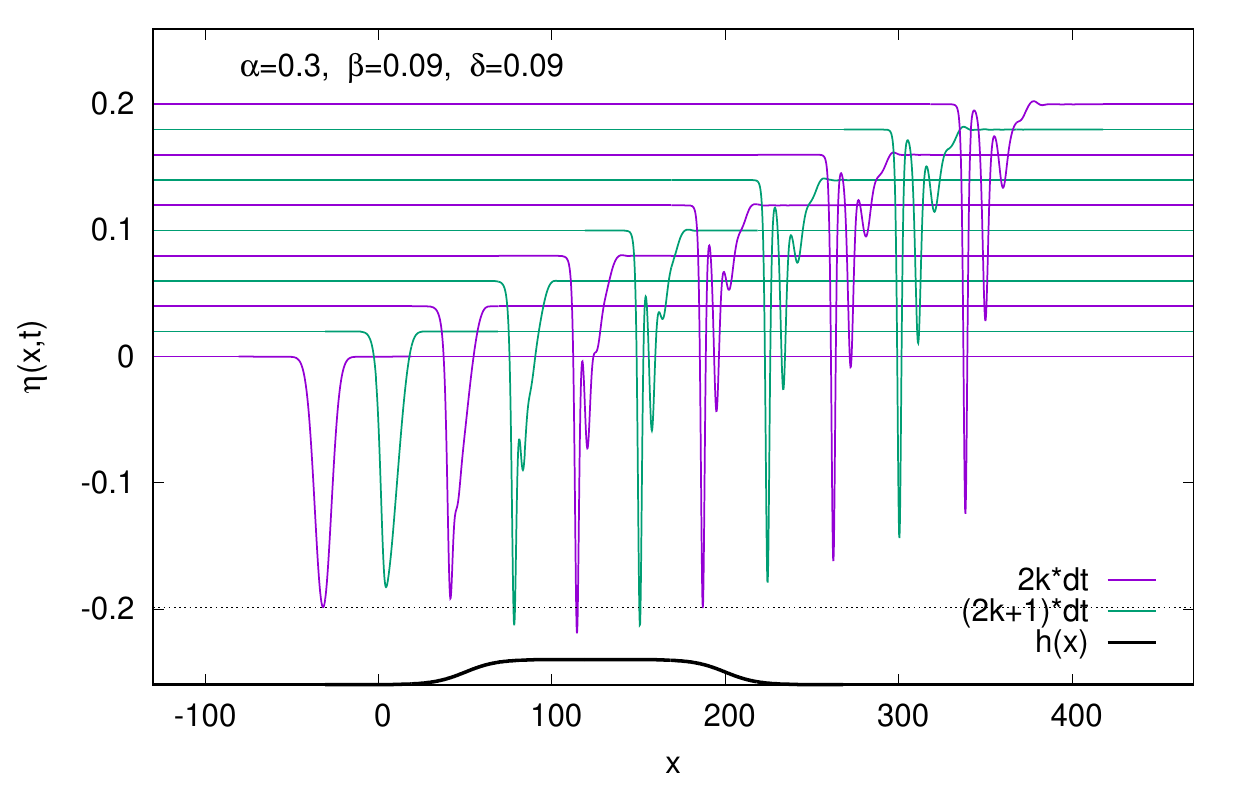}} \caption{Time evolution obtained according to the Gardner equation (\ref{GardH}) with $\tau=1$. Initial Gaussian profile with volume three times greater than that of the Gardner soliton, the same amplitude, and velocity. $dt=32$.} \label{a3bd09t1_Gs3a.}\end{center} \end{figure}

\begin{figure}[bht] \begin{center} \resizebox{0.9\columnwidth}{!}{\includegraphics{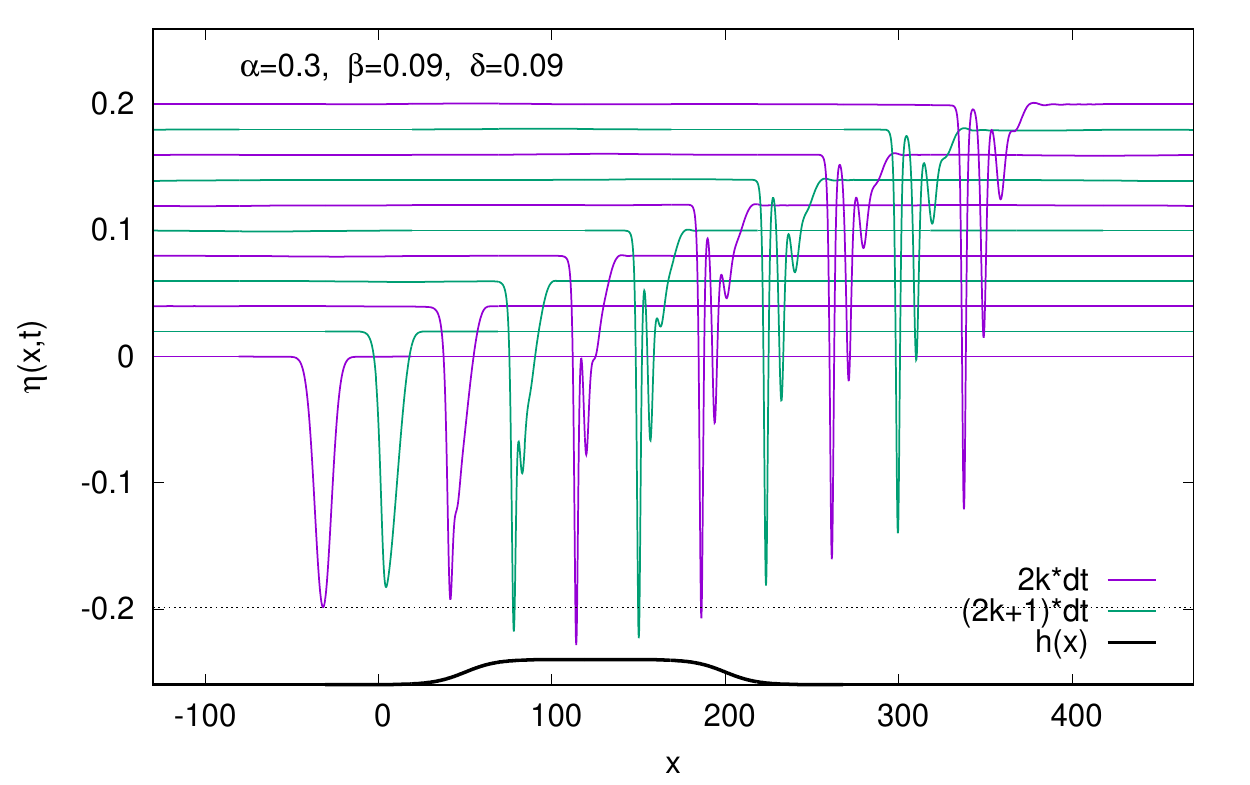}} \caption{Time evolution obtained according to Boussinesq's equations (\ref{bbu3})-(\ref{bbu4}) with $\tau=1$. Initial Gaussian profile with volume three times greater than that of the Gardner soliton, the same amplitude, and velocity.  $dt=32$.} 
\label{a3bd09t1_GBs3a.}\end{center} \end{figure}

\begin{figure}[bht] \begin{center}
 \resizebox{0.9\columnwidth}{!}{\includegraphics{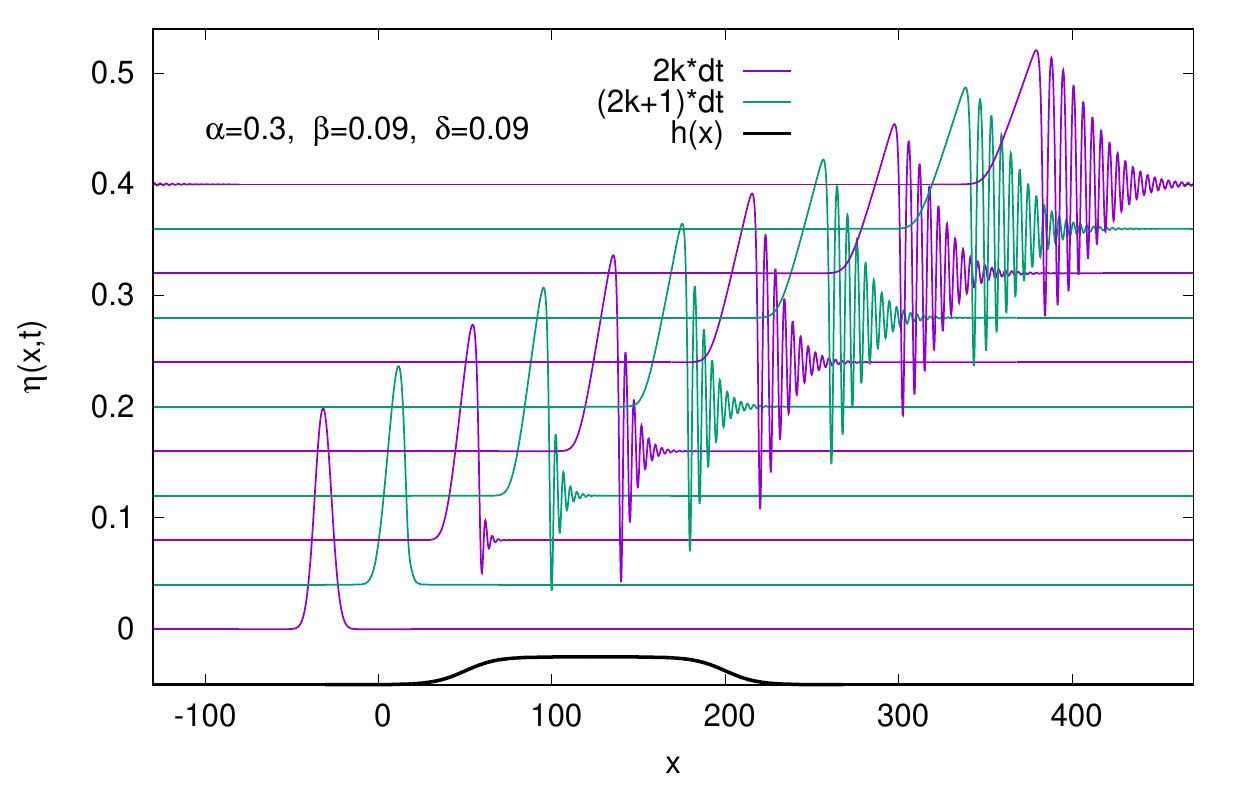}} \caption{Time evolution obtained according to the Gardner equation (\ref{GardH}) with $\tau=1$. Initial Gaussian profile, representing an elevation, with the triple volume of the Gardner soliton, but the same amplitude, and velocity.  $dt=32$.} 
\label{a3bd09t1_Gi3.}\end{center} \end{figure}

\begin{figure}[bht] \begin{center} \resizebox{0.9\columnwidth}{!}{\includegraphics{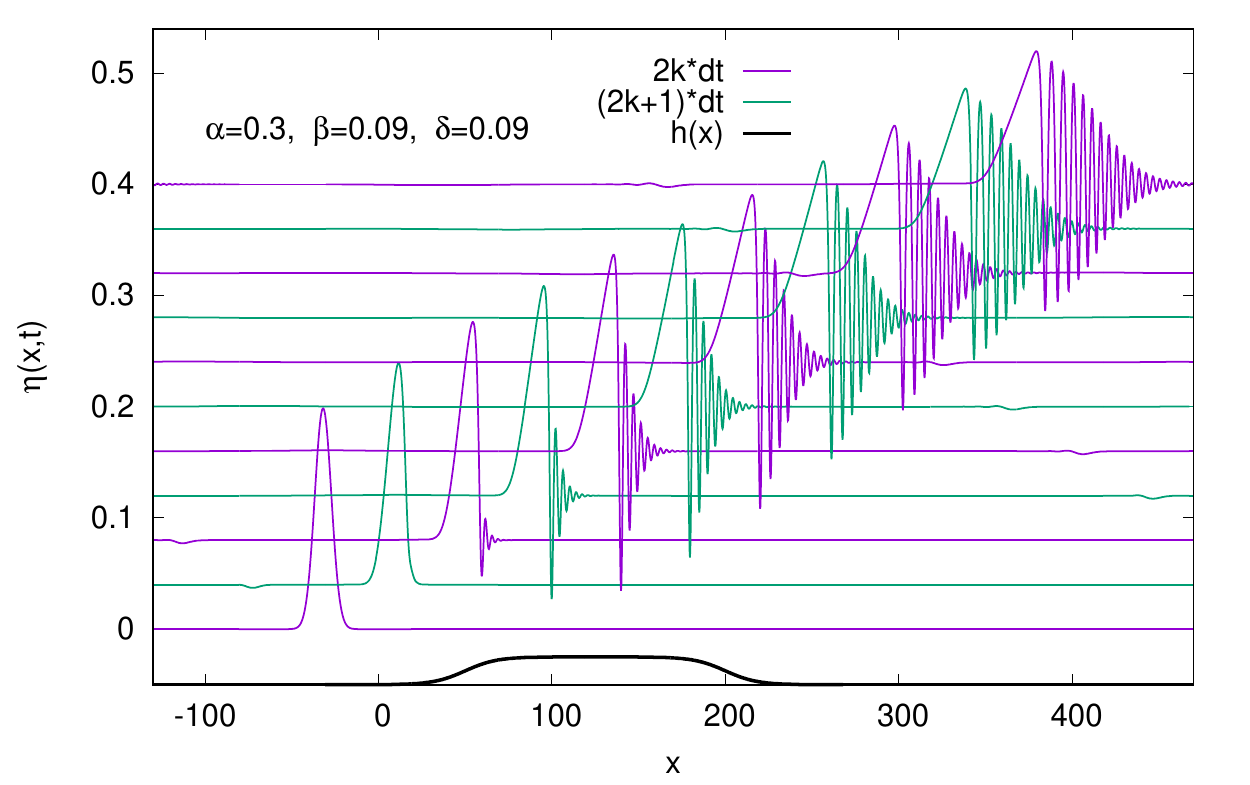}} \caption{Time evolution obtained according to Boussinesq's equations (\ref{bbu3})-(\ref{bbu4}) with $\tau=1$. Initial Gaussian profile, representing an elevation, with the the triple volume of Gardner soliton, but the same amplitude, and velocity.  $dt=32$.} 
\label{a3bd09t1_GBi3.}\end{center} \end{figure}

\section{Conclusions} \label{concl}

In all considered cases for the uneven bottom, the nonlinear wave equations (\ref{kdvD}), (\ref{nkdv2d}), (\ref{5kdvQ}), and (\ref{GardH})  
are non-integrable. Therefore the influence of the bottom variations on surface waves has to  be analyzed numerically. One must remember that the validity of the derived equations is limited to parameters $\alpha,\beta,\delta$ that are small enough. 

The main property of the results is the fact that the influence of the uneven bottom on the surface wave $\eta(x,t)$ obtained from the Boussinesq equations is always substantially greater than that obtained from single KdV-type wave equations. It is worth emphasizing that using the  Boussinesq equations does not need any conditions imposed on the form of the bottom function, whereas the compatibility condition, necessary for the existence of single KdV-type wave equations, requires $\frac{d^{2} h}{dx^{2}}=0$. 

The results of all simulations, performed according to the Boussinesq equations reveal the fact that the relative changes of $w(x,t)$ functions are substantially more prominent than that of $\eta(x,t)$ functions.

In all cases discussed above, when the initial conditions were chosen in the form of soliton solutions to particular wave equations, the wave profiles appear extremely resistant to disturbances introduced by varying bottom.

\end{document}